\documentclass[aps,superscriptaddress,reprint]{revtex4-1}

\usepackage[utf8]{inputenc}
\usepackage{xcolor}
\usepackage{pst-node}
\usepackage{graphicx}
\usepackage{amsmath, amssymb, amsthm}
\usepackage{mathtools}
\usepackage{hyperref}
\usepackage[shortlabels]{enumitem}
\usepackage{dsfont}
\usepackage{amsthm}
\usepackage{amssymb}
\usepackage{amstext}
\usepackage{amsfonts}
\usepackage{nicefrac}
\usepackage{extarrows} 
\usepackage{graphicx}
\usepackage{relsize}
\usepackage{ulem}

\newcommand{\R}{\mathbb R}

\newcommand{\calH}{{\mathcal H}}
\newcommand{\plus}{{ \!\, + \!\, }}
\newcommand{\FCI}{{\rm FCI}}
\newcommand{\MSE}{{\rm MSE}}
\newcommand{\RAS}{{\rm RAS}}
\newcommand{\CAS}{{\rm CAS}}
\newcommand{\RASX}{{\rm RAS-\!X}}
\newcommand{\fit}{{\rm fit}}
\newcommand{\xbar}{\overline{x}}
\newcommand{\ybar}{\overline{y}}
\newcommand{\Psitilde}{\tilde{\Psi}}
\newcommand{\Etilde}{\tilde{E}}

\newcommand{\ctilde}{\tilde{c}}

\begin{document}

\title{Predicting the FCI energy of large systems to chemical accuracy from restricted active space density matrix renormalization group calculations} 

\author{Gero Friesecke}%
 \email{gf@ma.tum.de}
\affiliation{%
  Department of Mathematics, Technical University of Munich, Germany
}%

\author{Gergely Barcza}
\affiliation{%
Strongly Correlated Systems Lend\"ulet Research Group,
Wigner Research Centre for Physics, H-1525, Budapest, Hungary
}%

\author{Örs Legeza}
\email{legeza.ors@wigner.hu }
\affiliation{%
Strongly Correlated Systems Lend\"ulet Research Group,
Wigner Research Centre for Physics, H-1525, Budapest, Hungary
}%
\affiliation{
Institute for Advanced Study,Technical University of Munich, Germany, Lichtenbergstrasse 2a, 85748 Garching, Germany
}

\date{\today}% It is always \today, today,

\begin{abstract}
\noindent \textbf{Abstract.} 
We theoretically derive and validate with large scale simulations a remarkably accurate power law scaling of errors for the restricted active space density matrix renormalization group (DMRG-RAS) method [arXiv:2111.06665] in electronic structure calculations. This yields a new extrapolation method, DMRG-RAS-X, which reaches chemical accuracy for strongly correlated systems such as the Chromium dimer, dicarbon up to a large cc-pVQZ basis,  and even a large chemical complex like the FeMoco
with significantly lower computational demands than previous methods. 
The method is free of empirical parameters, performed robustly and reliably in all examples we tested, and has the potential to become a vital alternative method for electronic structure calculations in quantum chemistry, 
and more generally for the computation of strong correlations in nuclear and condensed matter physics.

\end{abstract}

\maketitle

\section{Introduction}
\label{sec:intro}

In light of tremendous progress in the past decade in transition metal chemistry~\cite{Swart-2016,Khedkar-2021,Feldt-2022}, photosynthesis~\cite{Pantazis-2009,Sharma-2014,Vinyard-2017,Lubitz-2019}, single molecular magnets~\cite{Kerridge-2015,Spivak-2017,Gaggioli-2018}, and relativistic chemistry for compounds including heavy
elements~\cite{Trond-2011,Pyykk-2012,Reiher-2014a,Tecmer-2016,Wenjian-2020} the demand for a generally applicable method to efficiently treat strong electronic correlations and reveal solutions with chemical accuracy cannot be overemphasized. 
Although the main features of the electronic states are often characterized by the static correlations, contributions of an intractable number of  high energy excited configurations with small weights, i.e., dynamical effects, can be crucial for an accurate theoretical description in light of experimental data~\cite{Becke-2013,Benavides-2017}.

Quite recently, a cross-fertilization of the conventional restricted active space (RAS) 
scheme~\cite{Malmqvist-1990,Jensen-2006,Malmqvist-2008,Lischka-2018,Casanova-2022} with
the density matrix renormalization group method\cite{White-1992b,White-1999} via the
dynamically extended active space procedure~\cite{Legeza-2003b,Barcza-2011} has emerged as 
a new powerful method~\cite{Barcza-2022}
to capture both static and dynamic correlations, and 
numerical benchmarks on molecules with notorious multi-reference characters have revealed 
various advantages of the new method 
with respect to conventional approaches~\cite{Bartlett-2005,Bartlett-2007,Lischka-2018,Evangelista-2018,Pulay-2011}.
Furthermore, mapping of quantum lattice models to ab initio framework paves the road to attack challenging problems that are untractable by conventional approaches~\cite{Legeza-2018,Shapir-2019,Moca-2020}. The dramatic reduction of entanglement  
for higher dimensional lattice models via 
fermionic mode transformation~\cite{Krumnow-2016,Krumnow-2021} also makes the resulting ab intio problems 
excellent candidates for the DMRG-RAS method.

In this work, we present a theoretical analysis of the new method for the first time and introduce a new extrapolation procedure, free of empirical parameters and fully ab-initio, which reveals the ground state energy of systems with full Hilbert space dimensions up to
$2.48\times10^{31}$
within chemical accuracy (1 kcal/mole or 0.0016 a.u.). The principal insight is that the DMRG-RAS error exhibits stable power law scaling with respect to the CAS error, allowing reliable extrapolation. The scaling law is rigorously proved for a simplified model which captures key features of quantum chemical Hamiltonians, and numerically demonstrated for real molecules. We also show that DMRG-RAS is an embedding method, in the sense that when orbitals are partitioned into two subspaces, CAS and EXT, 
the correlations between them are calculated self-consistently, in contrast to other post-DMRG approaches~\cite{Piecuch-1996,Veis-2016,Leszczyk-2022,Kurashige-2011,Kurashige-2013,Sharma-2014b,Saitow-2013,Cheng-2022} which provide corrections on top of the DMRG wave function.
Furthermore, DMRG-RAS is
variational and the error exhibits monotone decay as the CAS-EXT split increases, unlike TCC where non-monotone behaviour is observed ~\cite{Faulstich-2019a,Faulstich-2019b}.

The unique features of the new method, dubbed DMRG-RAS-X, are demonstrated via large scale calculations for various strongly correlated molecules up to a full orbital space with more than hundred orbitals. This is achieved by utilizing
algorithmic progress developed in the past two decades based on concepts of quantum information theory~\cite{Szalay-2015a}. 
Therefore, our novel approach presented below, relying on a rigorous error scaling, can be applied to general systems and has the potential to become a widely used tool to target strongly correlated systems,
in particular multi-reference problems 
in quantum chemistry\cite{Cramer-2006,Kurashige-2009,Szalay-2015a,Chan-2012,Baiardi-2020,Cheng-2022}, 
nuclear structure theory~\cite{Dukelsky-2004,Legeza-2015,Tichai-2022} and condensed matter theory~\cite{Schollwock-2005,Orus-2014,Zhang-2017}.

The setup of the paper is as follows. In Secs.~\ref{sec:dmrg-ras} and \ref{sec:ref-ham} we present the theory of the DMRG-RAS method, 
while is Secs.~\ref{sec:error_weakly}
and \ref{sec:error_general} we derive the
error scaling for weakly interacting systems and general systems, respectively.
In Sec.~\ref{sec:extrapolation} we introduce our new extrapolation method and in Sec.~\ref{sec:numerical} benchmark results obtained by large scale DMRG-RAS-X calculations are presented.
Our work closes with main conclusions and future perspectives.

\section{The DMRG-RAS method}
\label{sec:dmrg-ras}

\begin{figure}
    \centering
    \includegraphics[width=0.5\textwidth]{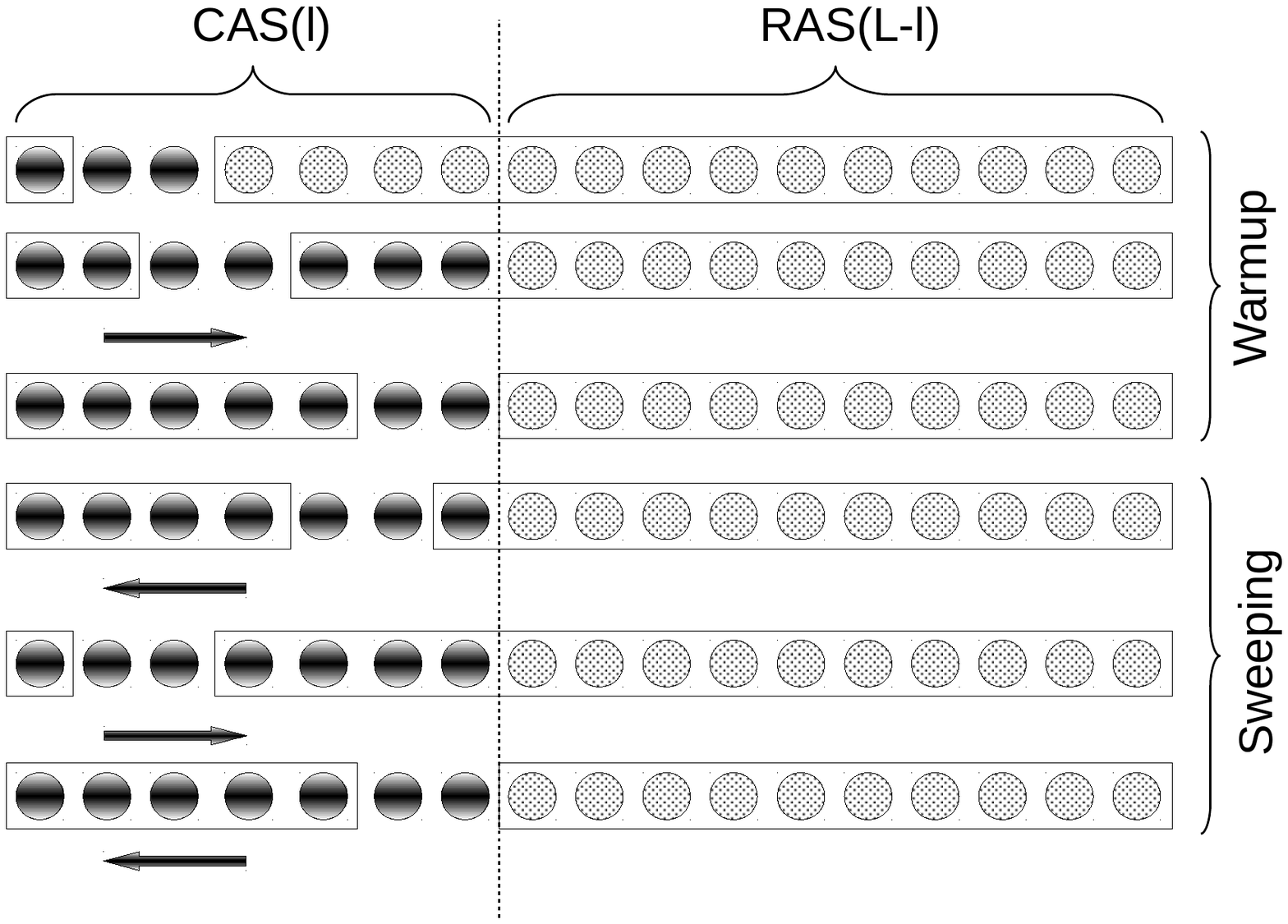}
    \vskip -6.0cm
    \caption{
    Partitioning of the orbitals into $\ell$ CAS orbitals and $L-\ell$ RAS orbitals, with $N/2\le \ell \le L$,
    in the DMRG-RAS method using the blocking structure introduced via the dynamically extended active space (DEAS) procedure. Filled circles stand for orbitals with four dimensional local Hilbert space, while orbital space built from dashed orbitals is restricted to an excitation threshold $k\le N$.
    Arrows indicate the DMRG sweeping procedure and the vertical line shows the turning point of the forward sweep as the RAS orbitals are treated as a single site.
    }
    \label{fig:dmrg-ras}
\end{figure}

The spinless single particle Hilbert space is spanned by $L$ orthonormal orbitals $\varphi_1,...,.\varphi_L \in L^2(\R^3)$, typically (but not necessarily) given by energy-ordered Hartree-Fock orbitals. In the DMRG-RAS method, one partitions the orbitals into $\ell$ CAS orbitals and $L-\ell$ RAS orbitals, with $N/2\le \ell \le L$ where $N$ is the number of electrons, and fixes an excitation threshold $k\le N$, see Figure~\ref{fig:dmrg-ras}. The DMRG related blocking structure via the dynamically extended active space (DEAS) procedure will be described in more detail below.
The $N$-electron Hilbert space for the DMRG-RAS method is then given by 
$$
  \calH(\ell,k) = \calH_{\rm CAS}(\ell) \bigoplus \calH_{\rm RAS}(L-\ell, k) 
$$
where the CAS Hilbert space is the full $N$-electron Hilbert space of the CAS orbitals,
$$
     \calH_{\rm CAS}(\ell) = \bigwedge_{i=1}^N span\{\varphi_1\uparrow,\varphi_1\downarrow,...,\varphi_\ell\uparrow,\varphi_\ell\downarrow\},
$$
and the RAS Hilbert space is spanned by all Slater determinants which are at least singly and at most $k$-fold excited with respect to some CAS Slater determinant,
\begin{align*}
     \calH_{\rm RAS}(L-\ell,k) & =  a^\dagger_{a_1\sigma_1}a_{j_1\sigma'_1}\calH_{\rm CAS} \bigoplus ... \\
     & \;\;\; \bigoplus 
                   a^\dagger_{a_1\sigma_1}...a^\dagger_{a_k\sigma_k}a_{j_1\sigma'_1}...a_{j_k\sigma'_k}\calH_{\rm CAS}.
\end{align*}
Here $a$ and $a^\dagger$ are the usual creation and annihilation operators, with CAS orbitals indexed by $j_i$, RAS orbitals indexed by $a_i$, and spin states indexed by $\sigma_i$, $\sigma'_i$.
Thus the method has two parameters, $\ell$ (number of CAS orbitals) and $k$ (RAS excitation threshold). While there is considerable freedom in choosing $\ell$, the standard choice for $k$ (and the one investigated in this paper) is $k=2$. 

Our interest is in the ground state energy $E^0=E^0(\ell,k)$ and the ground state $\Psi^0=\Psi^0(\ell,k)$ of the DMRG-RAS method obtained from the Rayleigh-Ritz principle, 
\begin{equation} \label{RR}
  E^0(\ell,k) = \min_{\Psi\in \calH(\ell,k) \, : \, \langle\Psi,\Psi\rangle=1} \langle \Psi, \, H \Psi\rangle,
\end{equation}
where $H$ is the (non-relativistic, Born-Oppenheimer) electronic Hamiltonian of the system, which can be written in second quantized form as
\begin{align} 
H &= \sum_{ij\sigma} t^{\phantom\dagger}_{ij\sigma}{a}^\dagger_{i\sigma}{a}^{\phantom\dagger}_{j\sigma}\nonumber\\ 
& +  \frac{1}{2} \sum_{ijkl \sigma\sigma^\prime}V^{\phantom\dagger}_{ijkl} {a}^\dagger_{i\sigma}{a}^\dagger_{j\sigma^\prime}{a}^{\phantom\dagger}_{k\sigma^\prime}{a}^{\phantom\dagger}_{l\sigma},
\label{ham}
\end{align}
where $t^{\phantom\dagger}_{ij}$ denotes the matrix elements of the one-particle Hamiltonian, which is
comprised of the kinetic energy and the external
electric potential of the nuclei, and  $V_{ijkl}^{\phantom\dagger}$ stands for 
stands for the matrix
elements of the electron repulsion operator.
Replacing the Hilbert space in \eqref{RR} by $\calH_{\rm CAS}(\ell)$ yields the CAS ground state energy and ground state $E^0_{\rm CAS}(\ell)$, $\Psi^0_{\rm CAS}(\ell)$.

For a closed-shell system, Hartree-Fock orbitals, and the smallest choice of $\ell$, i.e. $\ell=N/2$, the CAS Hilbert space is spanned by the Hartree-Fock determinant only. The CAS energy is then just the Hartree-Fock energy, and DMRG-RAS reduces to $k$-fold excited CI, i.e., when $k=2$, to CISD. When $\ell > N/2$, DMRG-RAS is a multi-reference method. 

We note also that 
for the largest (but typically unfeasible) choice $\ell=L$, the CAS energy $E^0_{\rm CAS}(L)$ is the full configuration interaction (FCI) energy, regardless of the choice of orbitals. Likewise, for the largest choice $k=N$ of the excitation threshold the energy $E^0(\ell,N)$ again gives the FCI energy, since $\calH(\ell,N)$ is the full $N$-electron Hilbert space. Our central goal is a quantitative understanding of how accurately $E^0(\ell,k)$ captures the FCI energy.

In the numerical simulations, the DMRG-RAS calculations will be performed to obtain $E^0(\ell,2)$ within a given error margin. 
In our implementation, we build on the 
dynamically extended active space (DEAS) procedure introduced two decades ago~\cite{Legeza-2003b}, where a multi-particle truncated Hilbert space is formed from a truncated set of one particle Hilbert spaces and DMRG right block auxiliary operators are calculated directly on that. A schematic picture of the partitioning of the 
$\ell$ CAS orbitals and $L-\ell$ RAS orbitals
into the DMRG blocks through the forward and backward sweeping procedure is given in Fig.~\ref{fig:dmrg-ras}.
For the warmup procedure, i.e., during the first forward sweep the right block is represented by the RAS space being equivalent to CISD for the CI-based truncated Hilbert space  (CI-DEAS)~\cite{Barcza-2011,Szalay-2015a} until the size of the left block reaches the desired value of $\ell-2$.
In all subsequent sweeps, the usual DMRG optimization steps are carried out for the CAS orbital space and in the current implementation the RAS space is left intact, i.e., it is simply attached to the DMRG chain. We note that a similar procedure has been introduced recently in Ref.~\cite{Larsson-2022}.
Using the dynamic block state selection (DBSS) approach~\cite{Legeza-2003a,Legeza-2004b} with tight error margin it can be guaranteed that the
error due to truncation in the DMRG procedure
remains much smaller than that due to the
Hilbert space truncation via a fixed RAS excitation threshold $k$.

\section{Theoretical analysis}
In this section, we provide the theoretical analysis of the DMRG-RAS method focusing on the scaling of the error as a function of $\ell$, and we also introduce a new extrapolation method to improve significantly the prediction of the full-CI energy. 

\subsection{The reference Hamiltonian}
\label{sec:ref-ham}

A key step in predicting the accuracy of $E^0(\ell,k)$ is a judicious partitioning of the full Hamiltonian into a reference Hamiltonian associated with the CAS energy and a remainder. We propose the following choice: 
\begin{eqnarray} H &=& H_0 + H' \mbox{ with } \label{H}\\
 H_0 &=& PHP + (E_0 + \Delta) Q  \label{H0} \\
 H'  &=& H - PHP - (E_0 + \Delta) Q \label{H'} 
\end{eqnarray}
where $P$ is the projector of $\calH$ onto the CAS Hilbert space $\calH_{\rm CAS}(\ell)$, $Q=I-P$ is the projector onto its orthogonal complement $\calH_{\rm RAS}(L-\ell,N)$ within the full $N$-electron Hilbert space $\calH(\ell,N)$, $E_0$ is the CAS ground state energy, i.e. 
\begin{align}
    E_0 = E^0_{\rm CAS}(\ell), \label{E0}
\end{align}
and $\Delta>0$ is a parameter to be chosen later. This partitioning has the following desirable features: 

(i) The CAS ground state energy is the ground state energy of $H_0$ (this is guaranteed by eqs.~\eqref{H0} and~\eqref{E0} and the positivity of $\Delta$);

(ii) The operator $H_0-E_0$ is invertible on the orthogonal complement of the ground state of $H_0$, yielding well-defined perturbation corrections at all orders; 

(iii) the first order perturbation correction $E^{(1)}=\langle\Psi_0|H'|\Psi_0\rangle$ vanishes regardless of the choice of the orbitals and parameters such as $\ell$ and $\Delta$; 

(iv) $H_0$ does not couple the CAS and RAS Hilbert spaces, with all the coupling contained in $H'$. 

The latter property is evident by re-writing
$$
   H - PHP = \!\!\! \underbrace{QHP}_{H_{\rm CAS\to RAS}}\! + \! \underbrace{PHQ}_{H_{\rm RAS\to CAS}} \! + \! \underbrace{QHQ}_{H_{\rm RAS\to RAS}} \!\!\! 
$$
(where the first term maps CAS to RAS, the second one, RAS to CAS, and the last one, RAS to itself), and makes transparent that DMRG-RAS can be considered an {\it embedding method}. 

The above partitioning is by no means the only one that achieves properties (i)--(iv), but it is perhaps the simplest, and allows to accurately predict the FCI energy, as shown below.

\subsection{Error scaling for weakly interacting systems}
\label{sec:error_weakly}

It is instructive to first discuss the case when $H'$ is small. So let us introduce a coupling constant $\lambda>0$ and look at the ground state energy $E_\lambda(\ell,k)$ of 
\begin{equation} 
\label{Hweak}
H_0+\lambda H'
\end{equation}
on $\calH(\ell,k)$ for small $\lambda$. We focus on the standard choice $k=2$. This energy shall be compared to the FCI energy $E^{\FCI}_\lambda = E_\lambda(\ell,N)$. 

Recall the reduced resolvent $R_0$ of $H_0$, 
$$
 R_0 = \begin{cases} 0 & \mbox{on the ground state of }H_0 \\  
 (H_0-E_0)^{-1} & \mbox{on (ground state)}^\perp 
 \end{cases}
$$
where (ground state)$^\perp$ denotes the orthogonal complement of the ground state within the full Hilbert space $\calH(\ell,N)$. In theoretical discussions of perturbation theory, the reduced resolvent is often expressed in terms of the excited eigenvalues and eigenstates of $H_0$ as $R_0 = \sum_{n\neq 0} (E_n-E_0)^{-1}|\Psi_n\rangle \langle \Psi_n|$; but this information is neither needed for our purposes, nor available in practice in cases like ours when $H_0$ is a many-body operator.

By standard Rayleigh-Schr\"odinger perturbation theory (see e.g.~\cite{aszabo82_qchem}), assuming the ground state of $H_0$ is nondegenerate and denoting it by $\Psi_0$, 
\begin{equation} \label{FCI}
  E_\lambda^{\FCI} = E_0 + \lambda E^{(1)} + \lambda^2 E^{(2)} + \lambda^3 E^{(3)} + O(\lambda^4) 
\end{equation}
as $\lambda\to 0$, with 
\begin{eqnarray}
\!\!\! E^{(1)} =\!\! & &\!\! \langle \Psi_0 | H' |\Psi_0\rangle, \\
\!\!\! E^{(2)} =\!\! & & \!\! \langle \Psi_0|H'|\Psi^{(1)}\rangle \nonumber \\
=\!\! &-&\!\! \langle\Psi_0|H'R_0H'|\Psi_0\rangle,  \\
\!\!\! E^{(3)} =\!\! & &\!\! \langle\Psi_0|H'R_0(H'\! -\! E^{(1)})R_0H'|\Psi_0\rangle \nonumber \\ 
=\!\! & & \!\! \langle\Psi^{(1)}|H'\! - \! E^{(1)}|\Psi^{(1)}\rangle .
\end{eqnarray}
Here 
\begin{equation} \label{Psi1}
  \Psi^{(1)} = -R_0 H' \Psi_0
\end{equation}
is the first order wavefunction correction under the usual intermediate normalization $\langle\Psi_0|\Psi_\lambda^{\rm FCI}\rangle=1$, so that 
\begin{equation} \label{FCIwave}
\Psi_\lambda^{\FCI}=\Psi_0 + \lambda \Psi^{(1)} + O(\lambda^2) 
\end{equation}
as $\lambda\to 0$. 

We now specialize to the Hamiltonians $H_0$ and $H'$ from \eqref{H0}, \eqref{H'}. The key point is that in this case, the quantum state obtained by applying $H'$ to the CAS ground state belongs to the doubly excited RAS space, 
\begin{equation} \label{Psi'}
  \Psi' = H'\Psi_0 \in \calH_{\rm RAS}(L-\ell,2).
\end{equation}
This is because (i) $H$ is a two-body operator, so it maps $\Psi_0$ into $\calH_{\rm CAS}(\ell)\oplus \calH_{\rm RAS}(L-\ell,2)$, (ii) $H'\Psi_0=QH\Psi_0$ is orthogonal to $\calH_{\rm CAS}(\ell)$.
Together with the fact that $R_0$ is just given by $\Delta^{-1}$ times the identity on $\calH_{\rm RAS}(L-\ell,2)$, it follows that 
\begin{align} 
E^{(1)}& =0, \nonumber \\
\Psi^{(1)} &= - \Delta^{-1} \Psi'\in \calH_{\rm RAS}(L-\ell,2), \nonumber \\
E^{(2)} &=  - \Delta^{-1} ||\Psi'||^2, \nonumber \\
 E^{(3)} &= \langle \Psi^{(1)}|H'|\Psi^{(1)}\rangle = \Delta^{-2}
      \langle \Psi'|H'|\Psi'\rangle. 
 \label{formulae}
\end{align}
We are now in a position to investigate the DMRG-RAS energy with excitation threshold $k=2$, i.e., $E_\lambda(\ell,2)$. First, to quantify its improvement over the CAS energy $E_0$, we use that  $\Psi^{(1)}$ belongs to the corresponding RAS space, so that $(\Psi_0+\lambda\Psi^{(1)})/||\Psi_0+\lambda\Psi^{(1)}||$ is an admissible trial function in the variational principle for $E_\lambda(\ell,2)$.  This gives 
\begin{align} 
 E_\lambda(\ell,2) & \le \frac{\langle \Psi_0 \plus \lambda \Psi^{(1)}|H_0+\lambda H'|\Psi_0 \plus \lambda \Psi^{(1)}\rangle} {1 + \lambda^2||\Psi^{(1)}||^2} \label{upbdnaive} \\
 & = E_0 + \frac{\lambda E^{(1)} + \lambda^2 E^{(2)} + \lambda^3 E^{(3)}}{1 + \lambda^2||\Psi^{(1)}||^2}.
 \label{upbd}
\end{align}
Here the last expression follows by multiplying out the numerator in \eqref{upbdnaive}, re-writing the term  $\lambda^2\langle\Psi^{(1)}|H_0|\Psi^{(1)}\rangle$ as  $\lambda^2 E_0 ||\Psi^{(1)}||^2 + \lambda^2 \langle\Psi^{(1)}|H_0\! -\! E_0|\Psi^{(1)}\rangle$, and using that  
$$
 \langle \Psi^{(1)}|H_0-E_0|\Psi^{(1)}\rangle = - E^{(2)} 
$$
(since $(H_0-E_0)R_0$ is the identity on the orthogonal complement of $\Psi_0$, to which $H'\Psi_0$ belongs). 

By Taylor-expanding 1 over the denominator as $1 - \lambda^2||\Psi^{(1)}||^2 + O(\lambda^4)$, 
$$
E_\lambda(\ell,2) \le E_0 + \lambda E^{(1)} + \lambda^2E^{(2)} + \lambda^3 E^{(3)} + O(\lambda^4),
$$
matching the expansion of the FCI energy to $O(\lambda^4)$. Together with the trivial lower bound $E_\lambda(\ell,2)\ge E_\lambda^{\FCI}$ this yields the overall error scaling
\begin{equation} \label{err_weak}
 {\tt e_\lambda}^{\rm RAS} = E_\lambda(\ell,2) - E_\lambda^{\FCI} = O(\lambda^4) \mbox{ as }\lambda\to 0.
\end{equation}
On the other hand, since $E^{(1)}=0$ and  $E^{(2)}<0$, the CAS error satisfies
\begin{equation} \label{err_CAS}
  {\tt e}_\lambda^{\rm CAS} = E_0 - E_\lambda^{\FCI} = |E^{(2)}|\lambda^2+O(\lambda^3) = \Omega(\lambda^2) \mbox{ as }\lambda\to 0.
\end{equation}
Here the Landau symbol  $\Omega(\lambda^p)$ denotes any term which stays bounded from below by a positive constant times $\lambda^p$ (analogously to the more common Landau symbol $O(\lambda^p)$ which denotes any term bounded above by a positive constant times $\lambda^p$). Combining with \eqref{err_weak} gives the following scaling law which relates the CAS and DMRG-RAS error: 
\begin{equation} \label{err_quotient}
   {\tt e}_\lambda^{\rm RAS}  = O\Bigl( ({\tt e}_\lambda^{\rm CAS})^2 \Bigr) \mbox{ as }\lambda\to 0.
\end{equation}

\subsection{Error scaling for fully interacting systems}
\label{sec:error_general}

Predictions of Rayleigh-Schr\"odinger perturbation theory have to be viewed with some caution, here (eq.~\eqref{err_weak}) and in many other cases (Moller-Plesset perturbation theory, G\"orling-Levy perturbation theory, ...). 
The reason is that the perturbing operator $H'$, which describes the action of the original Hamiltonian on  omitted pieces of the Hilbert space and between kept and omitted pieces, is not small, in the sense of smallness of its matrix elements $\langle\Psi|H'|\Phi\rangle$ or its operator norm 
\begin{equation} \label{opnorm}
  ||H'|| = \max_{||\Psi||=1} ||H'\Psi||=\max_{||\Psi||=||\Phi||=1}|\langle \Psi |H'|\Phi\rangle|.
\end{equation}
But the small parameter $\lambda$ in classical RS perturbation theory is precisely the operator norm $||H'||$ (as seen from writing $H' = \lambda \tilde{H}'$ with the fixed operator $\tilde{H}' =  {H'}/||H'||$). To summarize: neglecting higher-order energy contributions is rigorously justified only when $||H'||$ is small; but this condition is not met in practice.

What {\it can} be made small in practice in the DMRG-RAS method, by increasing the number $\ell$ of CAS orbitals to a reasonable value, is the {\it action of the perturbation operator $H'$ on the CAS ground state}, 
\begin{equation} \label{opnorm_on_CASGS}
     ||H'\Psi_0||,
\end{equation}
or its variant 
\begin{equation} \label{opnormtilde}
   ||H'\Psitilde_0||
\end{equation}
where $\Psitilde_0$ is the dressed CAS ground state introduced below. These quantities measures the total size of the Slater determinants in the RAS Hilbert space which get activated by applying the full Hamiltonian to the (original or dressed) CAS ground state. 
(The actual $\ell$ needed to make these quantities small depends on the system. Of course, a weakly correlated system can be captured sufficiently well by a small number of active orbitals, whereas a significantly larger $\ell$ is required for strongly correlated systems; see the numerical results below.)  
We now investigate to what extent the scaling laws \eqref{err_weak}, \eqref{err_CAS}, \eqref{err_quotient} remain correct when only \eqref{opnorm_on_CASGS} or \eqref{opnormtilde} (instead of \eqref{opnorm}) is small.  
Thus we aim to estimate 
the CAS and DMRG-RAS error at the physical value $\lambda=1$ in \eqref{Hweak}, in terms of $||H'\Psi_0||$ or $||H'\Psitilde_0||$. 

{\bf Lower bound on the CAS error.} 
We start from the variational upper bound \eqref{upbd}, which is valid for any $\lambda$, and obtain using \eqref{formulae}:
\begin{align}
  E^0(\ell,2) - E_0
  &\le   \frac{E^{(2)}+E^{(3)}}{1+||\Psi^{(1)}||^2}  \label{basicest}
  \\
  & = \frac{
  -\Delta^{-1} ||\Psi'||^2 + \Delta^{-2} \langle \Psi'|H'|\Psi'\rangle}{1+\Delta^{-2}||\Psi'||^2} .  
  \nonumber
\end{align}
The Taylor expansion $1/(1+\Delta^{-2}||\Psi'||^2)=1 - \Delta^{-2}||\Psi'||^2 $ $+ O(||\Psi'||^4)$ and the trivial estimate $|\langle \Psi' | H' | \Psi'\rangle |\le $  $||\Psi'|| \, ||H'\Psi'||$
now give
\begin{align}
   E^0(\ell,2)-E_0 &\le -\Delta^{-1}||H'\Psi_0||^2  \nonumber \\
   &+O(||H'\Psi_0|| \, ||H'{}^2\Psi_0|| + ||H'\Psi_0||^4). \label{remove} 
\end{align}
Assuming that 
\begin{equation} \label{scassptn}
    ||H'{}^2\Psi_0|| \ll ||H'\Psi_0|| \; \mbox{ as }||H'\Psi_0||\to 0,
\end{equation}
this inequality says that the CAS error contains a quadratic term in  $||H'\Psi_0||$, and that the DMRG-RAS method removes it. More precisely, combining \eqref{basicest} with the trivial fact $E_{\rm FCI}\le E^0(\ell,2)$ yields the universal lower bound
\begin{align} 
   {\tt \epsilon}_{\rm CAS}(\ell) &= E_0 - E_{\rm FCI} \ge E_0 - E^0(\ell,2) \nonumber \\
   &\ge \frac{|E^{(2)}|-E^{(3)}}{1+||\Psi^{(1)}||^2}
   \label{CASlowbd}
\end{align}
and combining \eqref{remove}--\eqref{scassptn} with $E_{\rm FCI}\le E^0(\ell,2)$ gives the asymptotic lower bound
\begin{align}
   \epsilon_{\rm CAS}(\ell)  = \Omega(||H'\Psi_0||^2) \mbox{ as }||H'\Psi_0||^2\to 0. \label{err_CAS_fancy} 
\end{align}

{\bf Upper bound on the DMRG-RAS error.} To obtain such an upper bound, i.e. a lower bound on the full CI energy, we rely on two ingredients, a spectral gap assumption and the notion of dressed CAS ground state. 

{\it Spectral gap assumption.} We take $\Delta$ in eqs. \eqref{H0}--\eqref{H'} to be the gap between the pure RAS  and the pure CAS ground state energies, 
\begin{align}
 \Delta &=\mbox{lowest eigenvalue of }H_{\rm RAS\to RAS} (=QHQ) \nonumber \\
 & -\mbox{lowest eigenvalue of }H_{\rm CAS\to CAS} (=PHP), \label{Deltadef}
\end{align}
and make the assumption that
\begin{equation} \label{gapassptn}
   \Delta > 0.
\end{equation}
(Here the RAS energy is that in the full RAS space $\calH_{\rm RAS}(L-\ell,N)$, see the definition of $Q$ in section \ref{sec:ref-ham}.) This spectral gap assumption is just a minimal (and in practice never violated) restriction on the choice of CAS and RAS orbitals. The above choice of $\Delta$ is independent of (and expected to be much larger than) the spectral gap between ground and first excited state within the CAS, and moreover increases monotonically with $\ell$, thereby promoting smallness of perturbation contributions which scale as inverse powers of $\Delta$. 

{\it Dressed CAS ground state.} By this we mean the normalized projection of the full FCI ground state $\Psi_{\rm FCI}$ onto the CAS,
\begin{equation} \label{dressed}
   \Psitilde_0 = \frac{P\Psi_{\rm FCI}}{||P\Psi_{\rm FCI}||}.
\end{equation}
This state gives rise to the dressed CAS ground state energy
$$
   \Etilde_0 = \langle \Psitilde_0 | H_0 | \Psitilde_0\rangle 
$$
and dressed perturbation contributions $\tilde{E}^{(i)}$ and  $\tilde{\Psi}^{(i)}$ ($i\ge 1$), defined by replacing the CAS ground state $\Psi_0$ in \eqref{Psi'}--\eqref{formulae} by the dressed ground state $\tilde{\Psi}_0$. Similarly to any embedded quantum system, the dressed ground state and perturbation contributions reflect the entanglement with the environment.

{\it Lower bound on the FCI energy.} We claim that the FCI energy satisfies the following rigorous lower bound:
\begin{equation} \label{FCIlowbd}
   E_{\rm FCI} \ge \Etilde_0 + \Etilde_2 - 4 ||\Psitilde^{(1)}||^2(\Etilde_0-E_0).
\end{equation}
To not interrupt our analysis of the DMRG-RAS method, the proof is relegated to an appendix. 

{\it Matching upper bound on the DMRG-RAS energy.} A closely matching upper bound on the DMRG-RAS energy is obtained by using, instead of the bare first-order-corrected CAS ground state $(\Psi_0+\Psi^{(1)})/||\Psi_0+\Psi^{(1)}||$, the dressed first-order-corrected CAS ground state $(\Psitilde_0+\Psitilde^{(1)})/||\Psitilde_0 + \Psitilde^{(1)}||$ as a trial function in the variational principle for $E^0(\ell,2)$. Note that this trial function is admissible because, for the same reasons underlying eq.~\eqref{Psi'}, 
\begin{equation} \label{Psitilde'}
    H'\Psitilde_0 \in \calH_{\rm RAS}(L-\ell,2). 
\end{equation}
It follows analogously to \eqref{basicest} that
\begin{equation} \label{fancyest}
 E^0(\ell,2)
  \le  \Etilde_0 + \frac{\Etilde^{(2)}+\Etilde^{(3)}}{1+||\Psitilde^{(1)}||^2} .
\end{equation}

{\it DMRG-RAS error.} Combining \eqref{FCIlowbd} and \eqref{fancyest} gives
\begin{align}
    \epsilon_{\rm RAS}(\ell) &= E^0(\ell,2) - E_{\rm FCI} \nonumber \\
    & \le \mbox{[r.h.s. of \eqref{fancyest}]} - \mbox{[r.h.s. of \eqref{FCIlowbd}]} \nonumber \\ 
    & = \frac{\Etilde^{(3)} + ||\Psitilde^{(1)}||^2|\Etilde^{(2)}|}{1 + ||\Psitilde^{(1)}||^2} + 4 ||\Psitilde^{(1)}||^2 (\Etilde_0 - E_0).
    \label{RASupbd}
\end{align}
 Consequently 
\begin{align} 
   \epsilon_{\rm RAS}(\ell) = &O\Bigl(\Etilde^{(3)}+||H'\Psitilde_0||^4 + ||H'\Psitilde_0||^2(\Etilde_0-E_0)\Bigr)
   \nonumber \\ & \mbox{ as } H'\Psitilde_0\to 0. \label{err_fancy}
\end{align}
Let us discuss the meaning of these error bounds. 

(i) Eq.~\eqref{RASupbd} shows that the RAS error tends to zero if the action of the perturbation operator $H'$ on the dressed CAS ground state $\Psitilde_0$ does so. This is because the latter implies that the r.h.s. of \eqref{RASupbd} tends to zero. 

(ii) Note that for this conclusion {\it neither} smallness of $H'$ on the complement of the CAS {\it nor} smallness of CAS-complement coupling is required, just a spectral gap and smallness of $H'$ on a specific low-energy CAS state (which is expected to be close to the CAS ground state).

(iii) It turns out in exactly soluble model examples representative of quantum chemical systems (see the next section) that the estimate \eqref{err_fancy} is typically sharp (as is our lower bound \eqref{err_CAS_fancy}). Unlike predicted by standard perturbation theory, the DMRG-RAS method does not in general capture the term $\Etilde^{(3)}$ which is cubic in $H'$!

(iv) Since $H'\Psitilde_0$ belongs to $\calH_{\rm RAS}(L-\ell,2)$, computation of the error bound only requires norm and energy evaluation of CAS and RAS states.

(v) The bounds \eqref{err_CAS_fancy}, \eqref{err_fancy} do not immediately reveal the scaling of the RAS error with respect to $||H'\Psitilde_0||$ or with respect to the CAS error. This scaling turns out to be system-dependent -- exactly so in our model examples in the next section, and approximately so in our numerical results for real molecules in Section \ref{sec:numjust}.

\subsection{Ladder model}
We now exhibit a simplified model which captures key features of the Hamiltonian \eqref{ham} and the hierarchical structure of the CAS and RAS Hilbert spaces, and sheds light on the quantitative relation between the CAS and RAS error. Select a sequence of normalized quantum states $\Psi_1$, $\Psi_2$, $\Psi_3$, ... such that 
$\Psi_1$ belongs to $\calH_{\rm CAS}(\ell_0)$
%each state only contains Slater determinants that are at least singly and at most doubly excited with respect to the Slater determinants of the previous state. That is,  $\Psi_1\in\calH_{\rm CAS}(\ell_0)$, $\Psi_2\in \calH_{RAS}(L-\ell_0,2)\cap \calH_{CAS}(\ell_0+2)$, $\Psi_3\in \calH_{RAS}(L-(\ell_0+2),2)\cap \calH_{CAS}(\ell_0+4)$, etc.
and $\Psi_j$ only contains Slater determinants which are $2(j\!-\! 1)$-fold excited with respect to $\calH_{\rm CAS}(\ell_0)$. 
Because of the two-body structure of the Hamiltonian \eqref{ham}, the resulting matrix restricted to the span of the $\Psi_j$ has the form 
\begin{equation} \label{eq:genham}
 \begin{pmatrix} h & {\rm v} & 0 & 0 & \cdots \\[-1mm]
 {\rm v} & h' & {\rm v'} & 0 & \ddots  \\[-1mm]
0 & {\rm v'} & h'' & {\rm v}'' & \ddots \\[-1mm]
0 & 0 & {\rm v''} & h''' & \ddots \\[-1mm]
 \vdots &  \ddots & \ddots & \ddots & \ddots 
 \end{pmatrix},
\end{equation}
where $h=\langle \Psi_1|H|\Psi_1\rangle$, ${\rm v}=\langle\Psi_1|H|\Psi_2\rangle$, $h'=\langle \Psi_2|H|\Psi_2\rangle$ etc. 
Typically the diagonal elements will increase and the off-diagonal elements will be of similar order of magnitude. We now make the following simplifying assumptions: 

(i) the diagonal elements increase linearly 

(ii) the off-diagonal elements are constant. 

\begin{figure}[http!]
    \centering
    \includegraphics[width=0.48\textwidth]{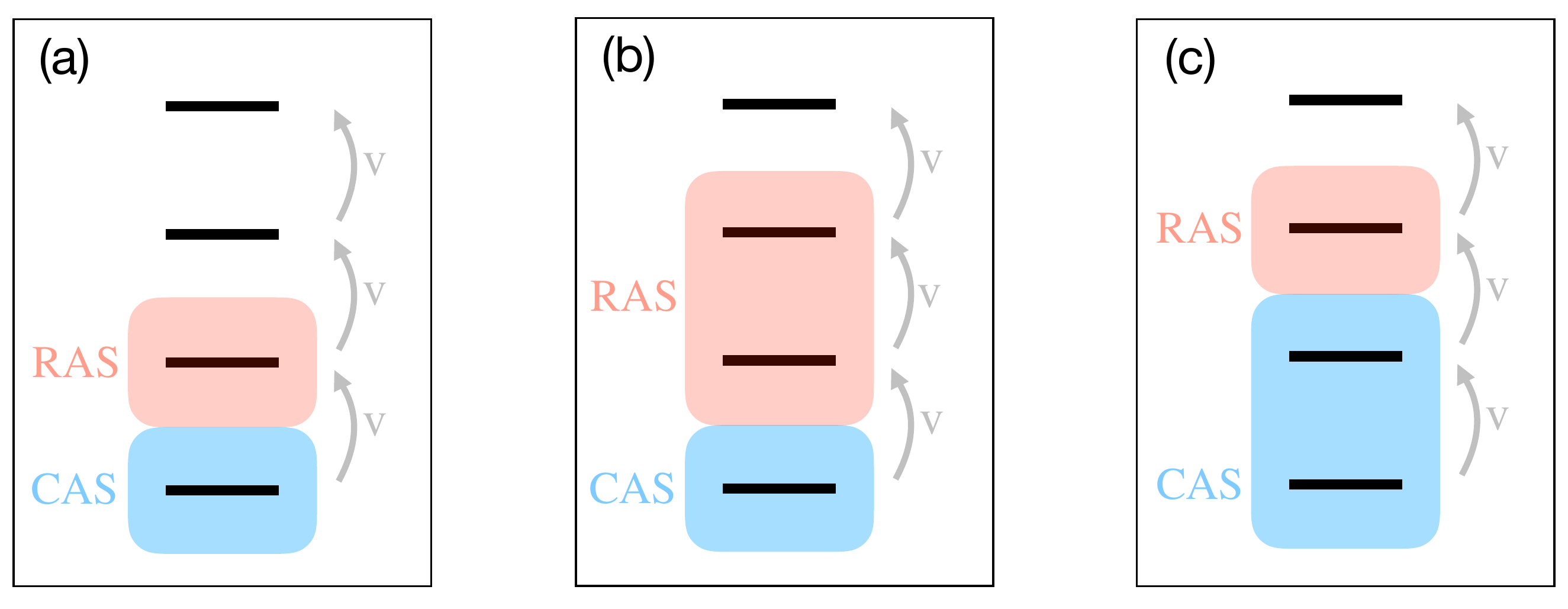} \\[-20mm]
    \includegraphics[width=0.44\textwidth]{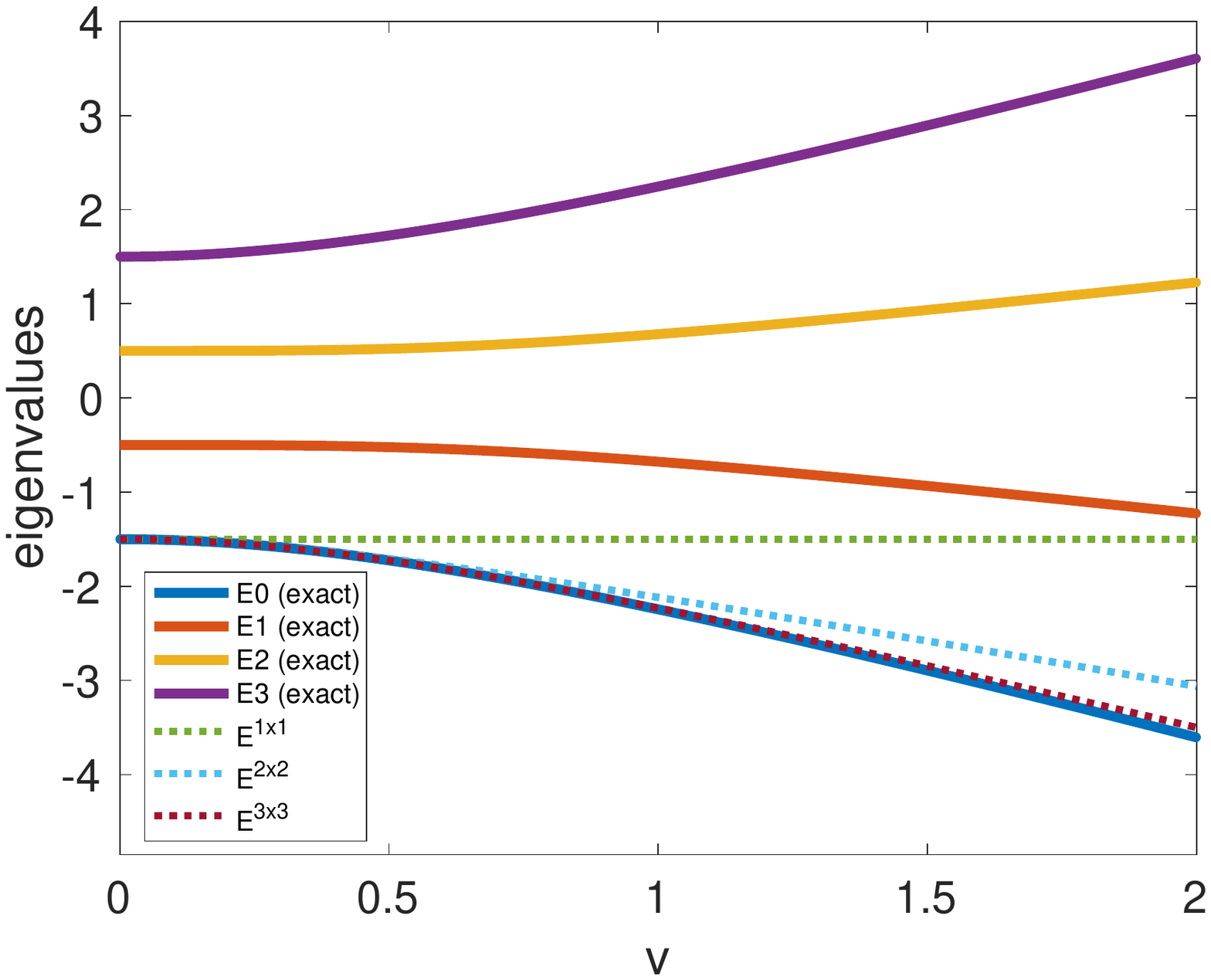}
    \vspace*{-23mm}
    \caption{Top: 4-state ladder model and possible choices of CAS and RAS. Bottom: Spectrum of the ladder model at fixed gap $g=1$ as a function of the interaction strength ${\rm v}$ (solid lines). The  ground state eigenvalues $E^{k\times k}$ of the top left $k\!\times\! k$ blocks of the Hamiltonian are also shown (dotted lines); these correspond to the CAS and RAS energies, with $k$ being the CAS dimension
    %(i.e. $1$ for (a) and (b) and $2$ for (c)) 
    respectively the sum of the CAS and RAS dimensions.}
    %(i.e. $2$ for (a) and $3$ for (b) and (c))
    \label{fig:ladder}
\end{figure}

\noindent
This yields the model Hamiltonian 
\begin{equation} \label{ladder}
H = \begin{pmatrix} h & {\rm v} & 0 & \cdots \\[-1mm]
                    {\rm v} & h\!+\! g & {\rm v} & \ddots  \\[-1mm]
                    0 & {\rm v} & h\!+\! 2g & \ddots \\[-1mm]
                    \vdots & \ddots & \ddots & \ddots \end{pmatrix},
\end{equation}
with gap parameter $g$ and interaction strength ${\rm v}$. 
Further, for simplicity let us focus on the case of just 4 states, that is, %
\begin{equation} \label{ladder4}
   H = \begin{pmatrix} h & {\rm v} & 0 & 0 \\[1mm]
                    {\rm v} & h\!+\! g & {\rm v} & 0  \\[1mm]
                    0 & {\rm v} & h\!+\! 2g & {\rm v} \\[1mm]
                    0 & 0 & {\rm v} & h\!+\! 3g
\end{pmatrix},
\end{equation}
which is exactly soluble (see below). 
We propose to call the model \eqref{ladder} the \textit{ladder model}. 
See Figure \ref{fig:ladder}. 
\vspace*{4mm}

In the 4-level case, there are exactly three possible choices of CAS and RAS, as depicted in the figure: (a) 1d CAS, 1d RAS, (b) 1d CAS, 2d RAS, (c) 2d CAS, 1d RAS. The resulting CAS respectively RAS energies $E_{\rm CAS}$, $E_{\rm RAS}$ are given by the lowest eigenvalue $E^{k\times k}$ of some top left $k\times k$ block $H^{k\times k}$ of $H$, where  $k$ is the CAS dimension respectively the sum of the CAS and RAS dimensions. E.g., in case (a), $E_{\rm CAS} = E^{1\times 1}$ and $E_{\rm RAS}= E^{2\times 2}$.

We observe that, with $\Psi_0$ and $\Psitilde_0$ denoting the CAS ground state and the dressed CAS ground state as before, in all three cases (a), (b) and (c), at fixed gap $g$ we have
\begin{equation}
    H'\Psi_0 \to 0 \Longleftrightarrow 
    H'\Psitilde_0 \to 0 \Longleftrightarrow {\rm v}\to 0. \label{equiv}
\end{equation}
Thus the abstract quantitites \eqref{opnorm}, \eqref{opnormtilde} reduce to the parameter ${\rm v}$ in the ladder model. 
Indeed, in cases (a) and (b) we have $\Psi_0=\Psitilde_0=\Psi_1$ so that $||H'\Psi_0||=||H'\Psitilde_0||=|{\rm v}|$, and in case (c) the expressions for the CAS and dressed CAS ground state derived in the appendix imply that $||H'\Psi_0||=(E^{1\times 1}-E^{2\times 2})/\sqrt{1+c^2}$ and $||H'\Psitilde_0||=(E^{1\times 1}-E^{4\times 4})/\sqrt{1+\ctilde^2}$,  establishing \eqref{equiv}. 

Hence in the following we study the CAS and RAS error as a function of the interaction strength ${\rm v}$ at fixed gap $g$, and denote them by $\epsilon_{\rm CAS}({\rm v})$ and $\epsilon_{\rm RAS}({\rm v})$. 

\begin{figure}
    \centering
      \includegraphics[width=0.235\textwidth]{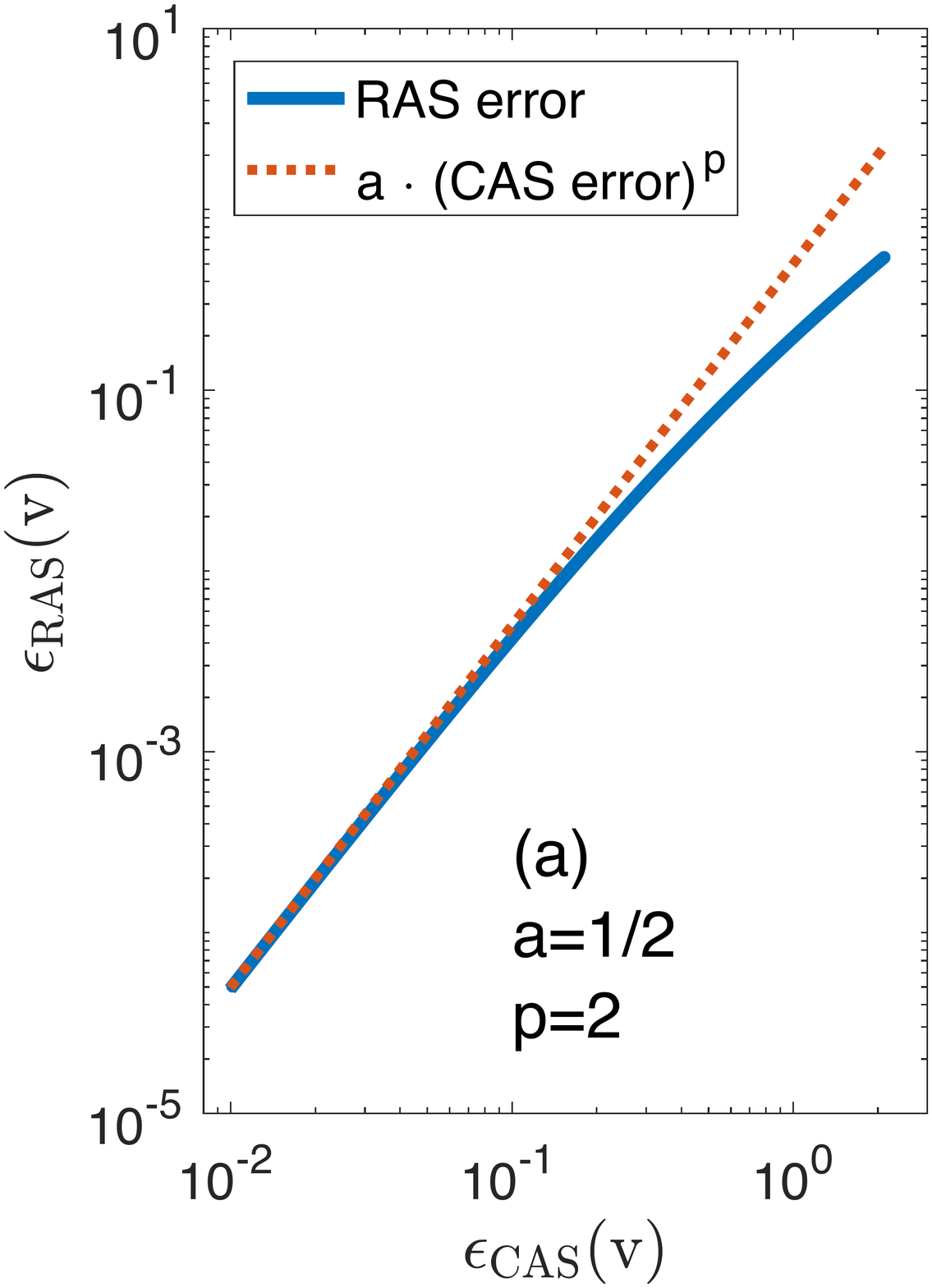}
   \includegraphics[width=0.235\textwidth]{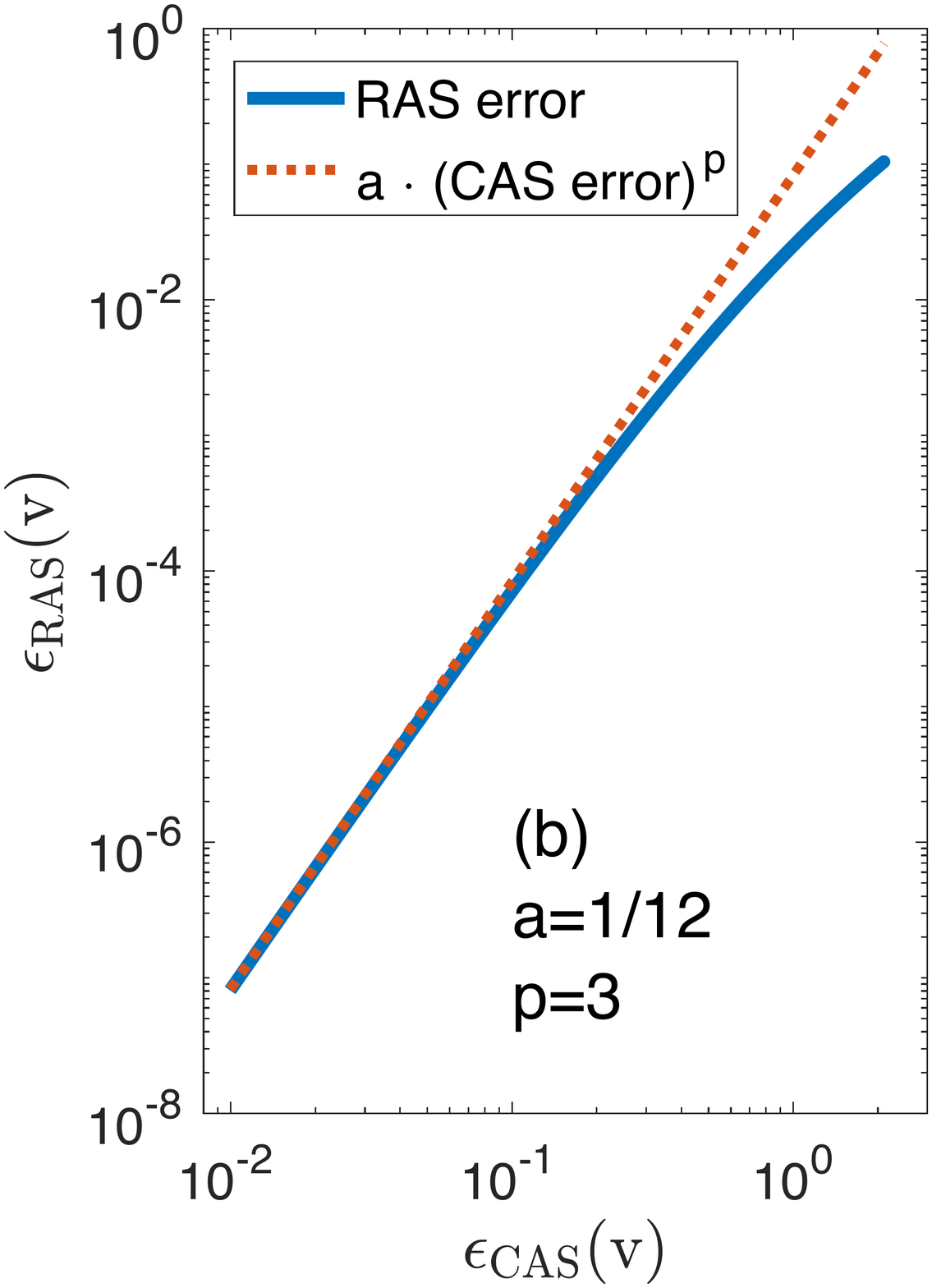} 
   \includegraphics[width=0.235\textwidth]{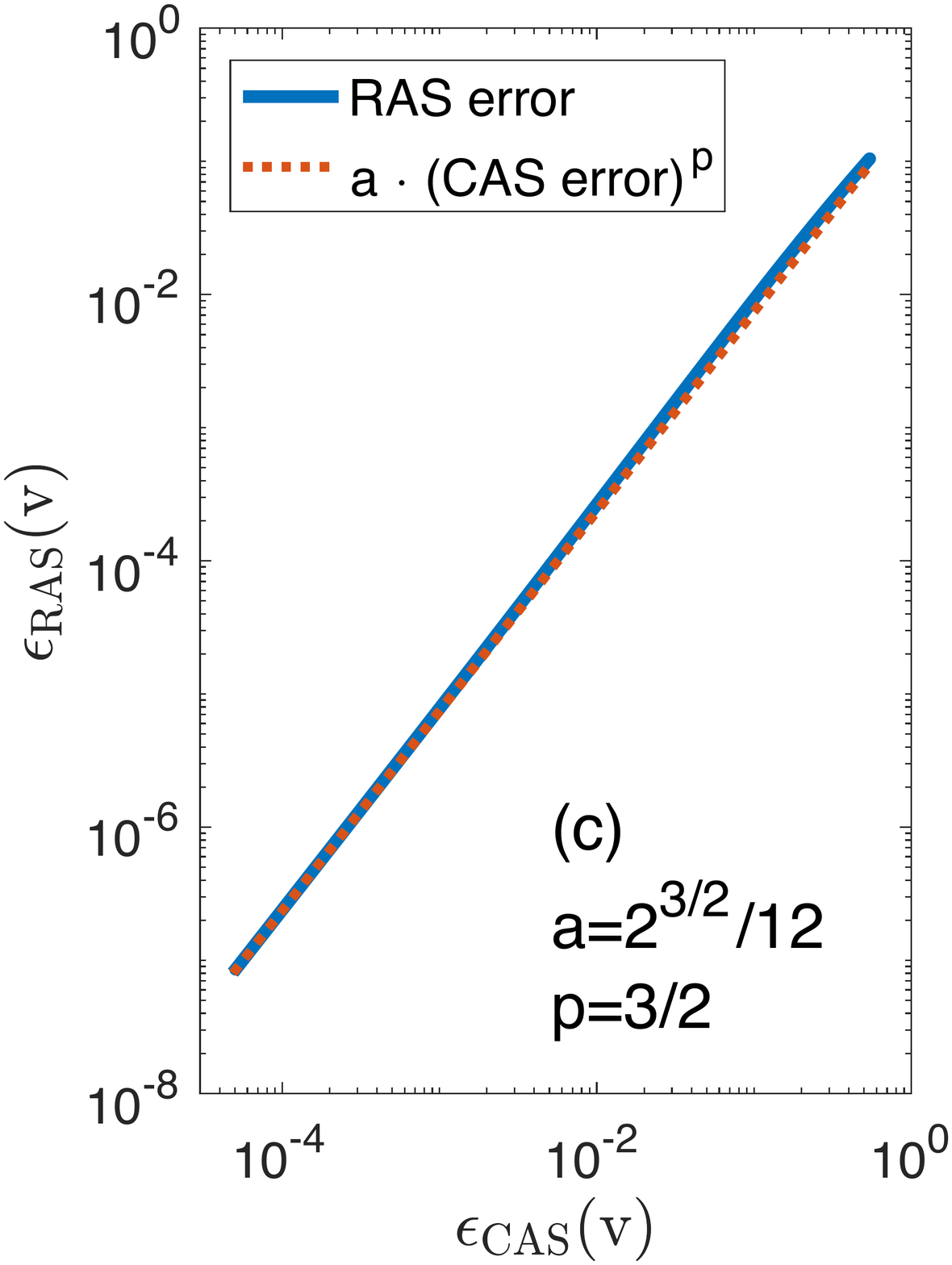}
    \caption{RAS error as a function of the CAS error shown on a double logarithmic scale for the ladder model with gap $g=1$ and interaction strength ${\rm v}$ varying from 0.1 to 2. The cases (a), (b), (c) correspond to the choices of CAS and RAS shown in Fig.~\ref{fig:ladder}. The dotted lines show the theoretical prediction from eqs.~\eqref{pred1}--\eqref{pred2}.}
    \label{fig:ladder_errors}
\end{figure}

We claim that the model exhibits an exact power law 
\begin{equation} \label{ladder_law}
   \epsilon_{\rm RAS}({\rm v}) \sim a \, \epsilon_{\rm CAS}({\rm v}){}^p \; \mbox{ as }{\rm v}\to 0,
\end{equation}
with non-universal exponent $p$ as given in Fig.~\ref{fig:ladder_errors}.

{\it Proof of the scaling law \eqref{ladder_law}.} The eigenvalues of the 4-state ladder model \eqref{ladder4}
can be found exactly, by solving the secular equation $\det(H-E\, {\rm Id)} = 0$. This is done by exploiting that the eigenvalues $E=E(h,g,v)$ satisfy $E(h,g,v)=E(\bar{h},g,v)+(h-\bar{h})$ and choosing $\bar{h}$ so as to make the Hamiltonian $H(\bar{h},g,v)$ traceless, $\bar{h}=-\tfrac{3}{2}g$, in which case the secular equation reduces to a quadratic equation for $E^2$ instead of a 4th order equation for $E$. This yields the following eigenvalues: 
\begin{align}
& h \, + \, \frac{3g}{2} \, \pm \, g\,\sqrt{\tfrac{5}{4} + \tfrac{3}{2}\bigl(\tfrac{{\rm v}}{g}\bigr)^2 \pm \sqrt{1 + 3\bigl(\tfrac{{\rm v}}{g}\bigr)^2+\tfrac{5}{4}\bigl(\tfrac{{\rm v}}{g}\bigr)^4}}.
\end{align}
For a plot of the eigenvalues at fixed gap $g=1$ as a function of the interaction strength ${\rm v}$ see Figure \ref{fig:ladder}. 

The required eigenvalues of the top left $k\times k$ block of $H$ can also be found exactly, trivially so for $k=1$ and $2$, and for $k=3$ by making the $3\times 3$ block traceless, in which case the middle eigenvalue of the block must be zero, again reducing the secular equation to just a quadratic equation. The result is
\begin{align}
& E^{1\times 1} = h \label{en1} \\[1.5mm]
& E^{2\times 2} = h \, + \, \frac{g}{2} \, - \, \frac{g}{2}\, \sqrt{1 + 4\bigl(\tfrac{{\rm v}}{g}\bigr)^2 } \label{en2} \\[0.25mm]
& E^{3\times 3} = h \, + \, g \, - \, g \, \sqrt{1 + 2\bigl(\tfrac{{\rm v}}{g}\bigr)^2} \label{en3} \\
& E^{4\times 4} = 
h \, + \, \frac{3g}{2} \, - \, g\,\sqrt{\tfrac{5}{4} + \tfrac{3}{2}\bigl(\tfrac{{\rm v}}{g}\bigr)^2 + \sqrt{1 + 3\bigl(\tfrac{{\rm v}}{g}\bigr)^2+\tfrac{5}{4}\bigl(\tfrac{{\rm v}}{g}\bigr)^4}} \label{en4}
\end{align}
Taylor-expanding in the relative interaction strength $\tfrac{\rm v}{g}$ using $\sqrt{1+z} = 1 + \tfrac{1}{2}z - \tfrac{1}{8}z^2 + \tfrac{1}{16}z^3 + O(z^4)$ yields 
\begin{align*}
& E^{2\times 2} \!= h  -  g \bigl(\tfrac{{\rm v}}{g}\bigr)^2 + g \bigl(\tfrac{{\rm v}}{g}\bigr)^4 + O(\bigl(\tfrac{{\rm v}}{g}\bigr)^6) \\
& E^{3\times 3} \!= h  -  g \bigl(\tfrac{{\rm v}}{g}\bigr)^2 + \tfrac{1}{2} g \bigl(\tfrac{{\rm v}}{g}\bigr)^4 - \tfrac{1}{2} g \bigl(\tfrac{{\rm v}}{g}\bigr)^6 + O(\bigl(\tfrac{{\rm v}}{g}\bigr)^8) \\
& E^{4\times 4}\! = 
h  -  g \bigl(\tfrac{{\rm v}}{g}\bigr)^2 + \tfrac{1}{2} g\bigl(\tfrac{{\rm v}}{g}\bigr)^4 - \tfrac{7}{12} g\bigl(\tfrac{{\rm v}}{g}\bigr)^6 + O(\bigl(\tfrac{{\rm v}}{g}\bigr)^8). 
\end{align*}
Hence the error $\epsilon^{k\times k}=E^{k\times k}- E^{4\times 4}$ satisfies, for $g=1$,
\begin{align*}
& \epsilon^{1\times 1} = {\rm v}^2 + O({\rm v}^4) \\
& \epsilon^{2\times 2} = \tfrac{1}{2}{\rm v}^4 + O({\rm v}^6) \\
& \epsilon^{3\times 3} = \tfrac{1}{12}{\rm v}^6 + O({\rm v}^8)
\end{align*}
as ${\rm v}\to 0$, 
and consequently 
\begin{align}
& \epsilon^{2\times 2} \sim \tfrac{1}{2}\bigl(\epsilon^{1\times 1}\bigr)^2, \label{pred1} \\
& \epsilon^{3\times 3} \sim \tfrac{1}{12} \bigl(\epsilon^{1\times 1}\bigr)^3 \sim \tfrac{2^{3/2}}{12} \bigl(\epsilon^{2\times 2}\bigr)^{3/2}. \label{pred2}
\end{align}
This establishes the scaling law 
\eqref{ladder_law}, with the values for $a$ and $p$ given in Fig.~\ref{fig:ladder_errors}.

For the infinite ladder model \eqref{ladder}, the above analysis suggests $\epsilon^{k\times k}\sim {\rm const} \cdot {\rm v}^{2k}$, which yields the following prediction for a $k$-dimensional CAS and an $r$-dimensional RAS:
\begin{equation}
   \epsilon_{\rm RAS}({\rm v}) \sim a \; \epsilon_{\rm CAS}({\rm v}){}^{\frac{k+r}{k}} \; \mbox{ as }{\rm v}\to 0.
\end{equation}
Thus a power law always holds, but the scaling exponent $p$ can be any rational number bigger than $1$.

{\it Stability of the power law.} The scaling law \eqref{ladder_law}, including the precise exponents from Figure \ref{fig:ladder_errors}, does not rely on the explicit expressions for the CAS, RAS, and FCI energies, but it can be recovered from the error bounds \eqref{err_CAS_fancy}, \eqref{err_fancy}. 
Incidentally, this demonstrates the high accuracy of these bounds in the asymptotic regime. In fact, these bounds only take into account that $H'(\calH_{\rm CAS})\subseteq\calH_{\rm RAS}$, so for the ladder Hamiltonian \eqref{ladder4} they are independent of the RAS dimension. (Refinements taking the latter into account are possible but not discussed here.) So our claim that these bounds recover  \eqref{ladder_law} and the exponents from Figure \ref{fig:ladder_errors} holds only for the cases (a) and (c) corresponding to a minimal RAS dimension. More precisely, in case (a) we find 
\begin{align}
  \mbox{r.h.s. of \eqref{CASlowbd} } &= \;\,{\rm v}^2 + O({\rm v}^4) \mbox{ as }{\rm v}\to 0, \label{app2-1} \\
  \mbox{r.h.s. of \eqref{RASupbd} } &= 2{\rm v}^4 + O({\rm v}^6) \mbox{ as }{\rm v}\to 0 \label{app2-2}
\end{align}
and in case (c) we find
\begin{align}
  \mbox{r.h.s. of \eqref{CASlowbd} } &= \tfrac{{\rm v}^4}{2} + O({\rm v}^6) \mbox{ as }{\rm v}\to 0, \label{app2-3} \\
  \mbox{r.h.s. of \eqref{RASupbd} } &= \tfrac{{\rm v}^6}{4} + O({\rm v}^8) \mbox{ as }{\rm v}\to 0, \label{app2-4}
\end{align}
exactly matching the scaling of the actual CAS and RAS errors and hence predicting the correct exponent $p$. A detailed derivation of \eqref{app2-1}--\eqref{app2-4} is given in an appendix. 

Let us close this section with 
some overall conclusions from our analysis of the ladder model. For general two-body Hamiltonians (as reflected by the form \eqref{eq:genham}) and general CAS and RAS spaces, we expect a scaling law
\begin{equation} \label{conclusion}
   \epsilon_{\rm RAS} \approx a \, \epsilon_{\rm CAS}^p.
\end{equation}
All exponents $p>1$ in this scaling law are possible; low exponents correspond to a good CAS, e.g. obtained by careful optimization of orbital space; and high exponents correspond to a poor CAS.

\subsection{New extrapolation method}
\label{sec:extrapolation}
 Suppose we have numerically calculated the CAS energy $E^0_{\rm CAS}(\ell)$ and the DMRG-RAS energy $E^0(\ell,2)$ for a few values of $\ell$.
The scaling law \eqref{conclusion} now gives some information about the (unknown) FCI energy, namely 
\begin{equation} \label{extra_p}
   E^0(\ell,2) - E^{\rm FCI} \; \approx \;  a \bigl(E^0_{\rm CAS}(\ell) - E^{\rm FCI}\bigr)^p \mbox{ for some }p>1.
\end{equation}
For careful numerical validation of this scaling law in real systems, with exponents $p$ differing from system to system as expected from theory, see the next section.

\color{black} 
Building on eq.~\eqref{extra_p}, we can predict the exponent $p$, the prefactor $a$ and the offset $E_{\rm FCI}$ from $E_{\rm CAS}^0(\ell)$ and $E^0(\ell,2)=E_{\rm RAS}(\ell)$.
This is achieved by minimizing the mean squared regression error of RAS versus CAS error in a log log plot,
\begin{equation} \label{MSE0}
    \MSE = \tfrac{1}{n} \sum_{\ell} \Bigl(y_\ell - (p\cdot x_\ell + \log a)\Bigr)^2 
\end{equation}
where $n$ is the number of datapoints and 
\begin{equation}
    x_\ell = \log\bigl(E_\CAS(\ell) - E_\FCI\bigr), \; \; 
    y_\ell = \log\bigl(E_\RAS(\ell)-E_\FCI\bigr).
\end{equation}
Vanishing of the regression error \eqref{MSE0} corresponds to exact validity of the scaling law \eqref{extra_p}. The minimization over the prefactor $a$ and the exponent $p$ can be carried out explicitly, yielding
\begin{equation} \label{MSE1}
   p_{\rm {\tiny \RASX}}=
   \frac{\sum_\ell (y_\ell - \ybar)(x_\ell-\xbar)}{\sum_\ell (x_\ell-\xbar)^2}, \;\; \log a_{\RASX} = \ybar - p_{\RASX} \xbar
\end{equation}
where $\xbar$, $\ybar$ are the average values of $x_\ell$ and $y_\ell$. This reduces the minimization of 
\eqref{MSE0} to a numerical minimization over the single free variable $E_{\rm FCI}$. The predicted FCI energy is then
$$
   E_{\RASX} = \underset{E^{\FCI}}{\rm arg \, min} \, \MSE, \;\; \MSE \mbox{ given by  \eqref{MSE0}--\eqref{MSE1}}.
$$

\section{Numerical results} \label{sec:numerical}

In this section we provide numerical justification for the theory presented in the previous section, focusing on the prediction given by Eq.~\ref{extra_p}. Here, let us remark that the overall error of the DMRG-RAS method stems from three main sources: (a) the split of the full orbital space to CAS and RAS parts; (b) the truncation of the RAS part based on $k$; and (c) the error of the DMRG used to approximate the CAS part.
The latter one can be controlled using the dynamic block state selection (DBBS) approach ~\cite{Legeza-2003b,Legeza-2004b} to stay below an a priori defined error threshold, $\chi$. The truncation on the RAS part is fixed by choosing $k=2$. Therefore, the main
question to be answered is the error dependence on $\ell$. We remark here  that the $\ell$-dependence hinges on a good choice of the CAS space and on the chosen basis ~\cite{Mate-2022, Barcza-2022}. Since orbitals lying close to the Fermi surface posses the largest one-orbital entropies~\cite{Legeza-2003b}, a selection based on orbital entropy together with keeping the almost fully occupied orbitals in the CAS space is an efficient protocol~\cite{Legeza-2003b,Stein-2016, Faulstich-2019b}.

Let us also note that a very accurate extrapolation procedure requires to perform CAS and DMRG-RAS calculations for various $\chi$ values for each $\ell$ in order to perform an extrapolation
to the DMRG truncation free limit, i.e, 
to obtain
$E^0(\ell) = \lim_{\chi\to 0}E^0_\chi(\ell)$
and
$E^0(\ell,k) = \lim_{\chi\to 0}E^0_\chi(\ell,k)$.
In this work, we have not performed this step since both
$\varepsilon_\CAS$ and $\varepsilon_\RAS$ were found to be larger by at least one order magnitude
than $\chi$ for the accessed $\ell$ values and for the systems studied below.

\subsection{Numerical justification of the scaling law} \label{sec:numjust}

First we investigate smaller systems with full-CI Hilbert space dimension not exceeding $10^{10}$, so that exact diagonalization can be carried out and the full-CI energy is available.  

\begin{figure}[h!]
 \includegraphics[width=0.48\textwidth]{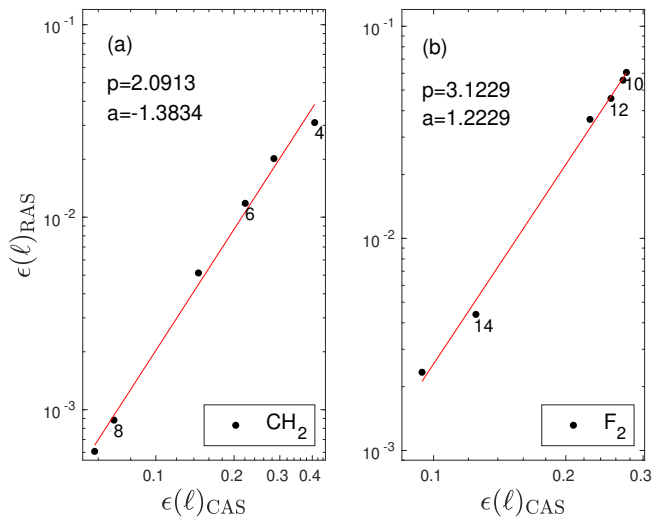}
  \caption{
(a) $\varepsilon(\ell)_{\rm RAS}$ as a function of $\varepsilon(\ell)_{\rm CAS}$ shown on a double logarithmic scale for the CH$_2$ molecule in the split valence basis corresponding to CAS(6,12).  The solid line is a fit according to Eq.~\ref{extra_p} and the obtained values for $p_\fit$ and $a_\fit$ are also shown. Labels next to data points stand for values of $\ell$.
(b) Similar to (a) but for the F$_2$ molecule at $d=2.68797{\rm a}_0$ using the basis of split valence orbitals 
corresponding to CAS(18,18). 
 \label{fig:cas_ras_ch2}}
\end{figure}

In Fig.~\ref{fig:cas_ras_ch2} we consider the CH$_2$ molecule. The absolute error of the DMRG-RAS ground state energy, 
$\varepsilon(\ell)_{\rm RAS}$,
is plotted as a function of the absolute error of
the CAS ground state energy, $\varepsilon(\ell)_{\rm CAS}$, on a double logarithmic scale, using the split valence basis. In the full orbital space this describes the correlation of 6 electrons on 12 (spatial) orbitals, i.e., CAS(6,12) with $\dim {\cal H}_\FCI\approx 1.3\times 10^{5}$. In order to acquire numerically exact data we have set the DMRG bond dimension to $D=16384$ and the residual error of the Davidson diagonalizaton method to $\varepsilon_{\rm Davidson}=10^{-11}$.
It is clearly seen in the figure that data points for increasing $\ell$ values fall on a line on the double logarithmic scale. This justifies the prediction of Eq.~\ref{extra_p}. In fact the fitted exponent comes out as $p_{\rm fit}=2.09$, a value that is very close to the value suggested by classical perturbation theory (eq.~\eqref{err_quotient}).

Next, we consider the F$_2$ molecule 
at $d=2.68797{\rm a}_0$
in the basis of split valence orbitals~\cite{Schafer1992}
corresponding to CAS(18,18) 
with $\dim {\cal H}_\FCI\approx 9\times 10^{9}$
which is almost the limit for exact diagonalization. Here, DMRG was performed using the DBSS procedure by setting $D_{\rm min}=2048$, $D_{\rm max}=15000$, and both the quantum information loss $\chi$ and 
$\varepsilon_{\rm Davidson}$
 to $10^{-6}$. The full-CI energy was taken from Ref.~\cite{Legeza-2003a}. Again data points lie along a line and the fitted value $p_\fit=3.12$ indicates that important corrections beyond perturbation theory are captured.  

\begin{figure}[h!]
\includegraphics[width=0.48\textwidth]{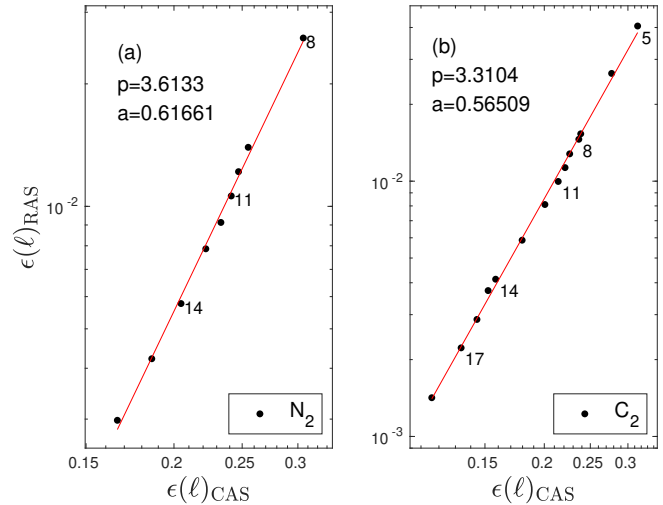}
  \caption{
(a) Similar to Fig.~\ref{fig:cas_ras_ch2} but for the N$_2$ 
molecule at $d=2.118{\rm a}_0$ in the cc-pVDZ basis  corresponding to CAS(14,28).
(b) Similar to (a) but for the C$_2$ 
molecule at $d=1.25\AA$ in the frozen core cc-pVTZ basis basis  corresponding to CAS(8,58). 
}
 \label{fig:cas_ras_n2}
\end{figure}

Turning towards much larger systems and with more pronounced correlations, i.e., to multi-reference problems, we show results for the N$_2$ molecule at inter molecular distance $d=2.118{\rm a}_0$ in the cc-pVDZ basis and for the C$_2$ molecule at $d=1.25\AA$, in the frozen core cc-pVTZ basis~\cite{Dunning-1989}, i.e., for CAS(14,28)
with $\dim {\cal H}_\FCI\approx 5.8\times 10^{12}$
and CAS(8,58) with $\dim {\cal H}_\FCI\approx 6.3\times 10^{11}$, respectively. 
In both cases, the same numerical settings 
have been applied as for F$_2$.
The numerically estimated full-CI ground state energy was taken from Ref.~\cite{Faulstich-2019b} and from
\cite{Barcza-2022}, respectively.
The data show near-perfect agreement with the behavior predicted from theory, and the slopes of the fits are $p_\fit=3.6$ and $3.3$, respectively.
Therefore, we can conclude that the power law scaling in eq.~\ref{extra_p} holds for realistic systems.

\subsection{Numerically predicting the FCI energy within chemical accuracy}

\begin{figure}[h!]
 \includegraphics[width=0.48\textwidth]{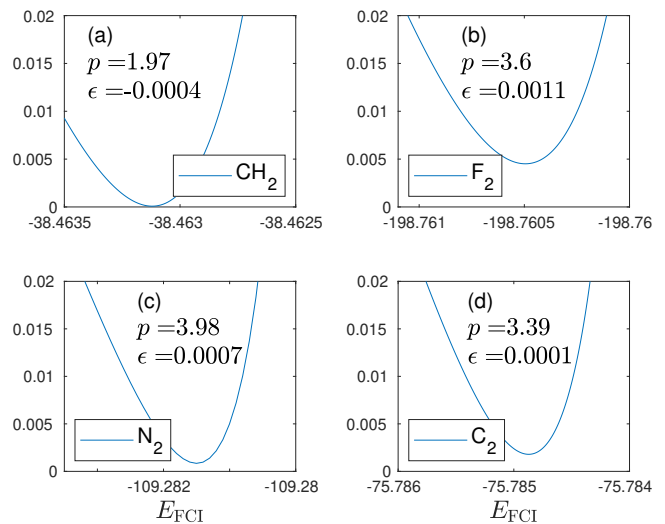}
  \caption{
(a),(b),(c) and (d) show the 
regression error, Eq.~\eqref{MSE0},
as a function of $E_{\FCI}$ for the CH$_2$, F$_2$, N$_2$, and for the C$_2$ molecules, respectively, at the predicted value $p_\RASX$ of the exponent. 
The latter, and the 
absolute error $\varepsilon_\RASX$
given by the predicted value of $E_{\FCI}$ 
minus the full-CI energy, 
are also included. 
 \label{fig:cas_ras_fit}}
\end{figure}

We now demonstrate that the new extrapolation method presented in Sec.~\ref{sec:extrapolation} by Eq.~\ref{extra_p}
can be applied to realistic systems, providing both the
ground state energy within chemical accuracy and an intrinsic error estimate for the DMRG-RAS calculation which does not require external reference data.
In Fig.~\ref{fig:cas_ras_fit} the result of this minimization procedure providing $p_{\RASX}$ and
$E_{\RASX}$
is presented for CH$_2$, F$_2$, N$_2$ and C$_2$. 
The predicted exponents are 
in all cases close to the fitted values obtained by using corresponding Full-CI reference energies (see Figs.~\ref{fig:cas_ras_ch2} and ~\ref{fig:cas_ras_n2}); in fact, in case of C$_2$, the difference is of the order of $10^{-2}$. More importantly, the extrapolation reduces the absolute error in the ground state energies significantly.
The lowest values of the absolute error in the ground state energy achieved at the largest values of $\ell$ for the CAS and RAS methods, together with extrapolated values, are summarized in Tab.~\ref{tab:error}. 
\begin{table}[t]
  \centering
\begin{tabular}{l|c|c|r|c|c}
\hline
 \hline
system$\;\;\,$ & $\;\;\varepsilon_{\CAS}\;\;$ & $\;\;\varepsilon_{\RAS}\;\;$ & $\varepsilon_{\RASX}$ & $\;\;\;\, L/\ell_{\rm max}\;\;\;\, $ & $\varepsilon_{\RASX}/\varepsilon_{\RAS}\!\!$  \\
  \hline
 F$_2$ &0.0941  &0.0023   &0.0011 &1.20 & 0.48 \\
 \hline
 CH$_2$ &0.0690  &0.0009  &-0.0004 &1.33& 0.29 \\
 \hline
 N$_2$ &0.1662 &0.0030 &0.0007 & 1.75 & 0.23 \\
 \hline
 C$_2$ &0.1159 &0.0014 &0.0001 & 3.22 & 0.07 \\
 \hline
 \hline
 \end{tabular}
\caption{Absolute error of the ground state energy for various systems,  
for the CAS, RAS, and RAS-X method. For the first two methods we used the largest $\ell$ values (i.e. the maximal number of CAS orbitals, $\ell_{\rm max}$) in Figures \ref{fig:cas_ras_ch2} and \ref{fig:cas_ras_n2}. Also shown: ratio $L/\ell_{\rm max}$ of total number of orbitals to maximal number of CAS orbitals, and ratio of RAS-X to RAS error. Note that the RAS-X method achieves chemical accuracy (1 kcal/mol or 0.0016 a.u.) in all cases.}
  \label{tab:error}%
\end{table}%
Our numerical data demonstrate that the absolute error is reduced by one to two orders of magnitude by the DMRG-RAS method with respect to the restricted CAS space solution, and it is further reduced by a factor of 2 to 15 via the new extrapolation procedure according to Eq.~\ref{extra_p}. This further reduction in error becomes monotonically better as the ratio $L/\ell_{\rm max}$ of total number of orbitals to maximal number of CAS orbitals increases, as is seen from the last two columns of the table. In addition, the difference between $E_{\RAS}$ and $E_{\RASX}$ 
provides an 
estimate the error of a DMRG-RAS calculation without reference to external data, which 
is a unique and in our opinion very important
feature among methods developed for electronic structure calculations.

As a final test we have repeated our analysis for stretched geometries of N$_2$, i.e, for bond lengths $d=2.700{\rm a}_0$ and $3.600{\rm a}_0$.  By increasing $d$ the multi reference character of the wave function, i.e., the contribution of dynamic correlations, becomes more pronounced.
In Table.~\ref{tab:n2}
the DMRG and CC energies obtained for the full orbital space, together with 
extrapolated full-CI energies, $E_{\RASX}$ and related exponents $p_{\RASX}$ using $\ell=8\ldots 16$
are summarized.
Values of $p_\fit$ obtained by direct fits
of $\varepsilon_\RAS$ vs $\varepsilon_\CAS$ 
using the full-CI reference energies,
as shown in Fig.~\ref{fig:cas_ras_n2}, are also included for comparison.
Our result confirms that the DMRG-RAS-X method provides very accurate energy values not only for equilibrium geometries but also in situations when dynamic correlations are even more pronounced.
\begin{table}[t]
  \centering
\begin{tabular}{l|c|c|r}
\hline
 \hline
  & $d=2.118{\rm a}_0$& $d=2.700{\rm a}_0$ & $d=3.600{\rm a}_0$ \\
 \hline
 $E_{\rm CCSD}$ &-109.26762 &-109.13166 &-108.92531  \\
 \hline
 $E_{\rm CCSDT}$ &-109.28032 &-109.15675 &-109.01408   \\
 \hline
 $E_{\rm CCDTQ}$ &-109.28194 &-109.16224 &-108.99752  \\
 \hline
 $E_{\RASX}$ &-109.2814 & -109.1634& -108.9980 \\
 \hline
 $p_{\RASX}$ & 3.98 &  3.45 &  3.23\\
 \hline
 $\varepsilon_\RASX$& 0.0007 & 0.0002 & 0.0001\\
 \hline
 $p_{\rm fit}$ & 3.61 &  3.34 &  3.20\\
 \hline
 $E_{\rm DMRG}$ & -109.282165  &-109.16359 &-108.99807  \\
 \hline
 \hline
 \end{tabular}
\caption{Full-CI ground state energies obtained by large-scale DMRG
calculations with $M_{\rm min}=1024$, $M_{\rm max}=10000$ and $\chi=10^{-6}$, together with
CC reference energies taken from Ref.~\cite{Faulstich-2019b} and predicted values $E_{\RASX}$, $p_{\RASX}$ and $\varepsilon_\RASX$ via the DMRG-RAS-X method using $\ell=8\ldots 16$ for various bond lengths for the N$_2$ dimer in the cc-pVDZ basis. The $p_{\rm fit}$ values correspond to direct fits of $\varepsilon_\RAS$ vs $\varepsilon_\CAS$ 
as shown in Fig.~\ref{fig:cas_ras_n2}
using the full-CI reference energies.}
  \label{tab:n2}%
\end{table}%

\subsection{Results for large basis sets and the stability of the DMRG-RAS-X method}

In this section, we present our results for 
larger systems and larger basis sets. 
For such systems the full-CI energy is not available, so our DMRG-RAS and DMRG-RAS-X results
can only be compared to calculations performed with conventional methods like CC or MRCI, or to reference data accessible in the literature.

\begin{figure}[h!]
 \includegraphics[width=0.48\textwidth]{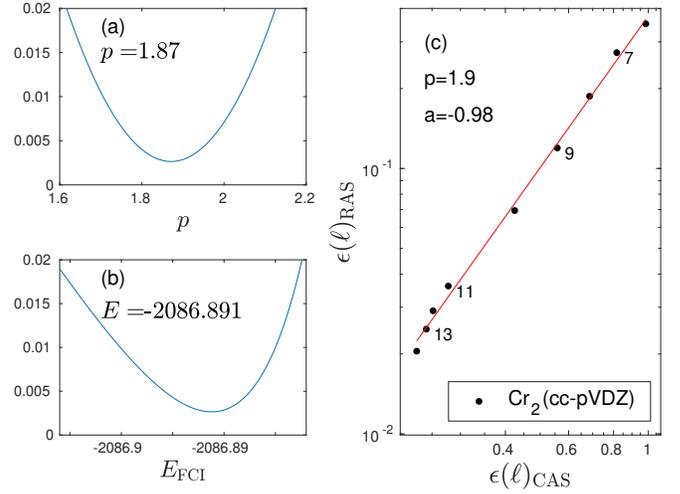}
  \caption{
Similar to Fig.~\ref{fig:cas_ras_ch2}, but for the chromium dimer at its equilibrium geometry, $d=1.6788 \AA$ in a natural orbital basis obtained from the cc-pVDZ atomic basis, corresponding to a full orbital space CAS(12,68). Here, first the extrapolation has been performed to obtain the predicted exponent $p_{\RASX}$ and  energy value $E_{\RASX}$ as shown in panel (a) and (b), and the predicted energy was used to get the curve presented in panel (c).
}
\label{fig:cas_ras_fit_cr2}
\end{figure}

First, we show our result for the notoriously strongly correlated chromium dimer which is subject to usual benchmark calculations even nowadays ~\cite{Kurashige-2011,Ma-2017,Sharma-2012,Veis-2016,Barcza-2022,Larsson-2022}. 
Our result for
Cr$_2$ in a natural orbital basis obtained from the cc-pVDZ atomic basis (see Ref.~\cite{Barcza-2022}) at its equilibrium geometry, $d=1.6788 \AA$, corresponding to a full orbital space CAS(12,68) with $\dim {\cal H}_\FCI\approx 5\times 10^{16}$, 
is shown in Fig.~\ref{fig:cas_ras_fit_cr2}.
In the numerical calculations we have used $M_{\rm min}=2048$, 
$M_{\rm max}=10000$, $\chi=10^{-6}$, $\varepsilon_{\rm Davidson}=10^{-6}$
and eleven sweeps.
Here, due to the large orbital space we do not have access to the full-CI energy, so we first perform our extrapolation scheme yielding the predicted parameters shown in Fig~\ref{fig:cas_ras_fit_cr2} (a) and (b).
Next, using the predicted full-CI energy, $E_{\RASX}=-2086.891$, we show in
Fig.~\ref{fig:cas_ras_fit_cr2} (c) that
the linear scaling on double logarithmic axes 
for different $\ell$ values
is recovered, as expected. Comparing our result to CCSD, CCSD(T), CCSDT, and CCSDTQ (E=-2086.7401, -2086.8785, -2086.8675, -2086.8689, respectively) we conclude that our data point $E^0(17,2)=-2086.8769$ is already below the CCSDTQ by $8\times10^{-3}$, while the extrapolated energy is between $E_\RASX=2086.884$ ($p_\RASX=2.06$) and $E_\RASX=2086.891$ ($p_\RASX=1.88$) using the first 12 or 14 data points in the fit,  leading to an error estimate of the order of $10^{-3}$.
We have excluded the last three data points from the fit ($\ell=15, 16, 17$) since the dynamically selected DMRG bond dimension has hit the upper threshold $M_{\rm max}$.

Next, we analyze the stability of the extrapolation 
with respect to the accuracy threshold imposed on DMRG. Lowering the accuracy, i.e., using $\chi=10^{-4}$, our data points for the largest $\ell$ values shifted upwards by the order of $10^{-3}$ and we obtained $E_\RASX=-2086.881$ with $p_\RASX=2.08$.
Increasing the value of $\chi$ further to $10^{-3}$
we obtained $E^0(14,2)=-2086.8599$ with $\max(M)\simeq 4000$ only, and the extrapolation
yielded $E_\RASX=-2086.873$ with $p_\RASX=2.02$.
Therefore, DMRG-RAS-X even with limited accuracy, i.e., with limited computational demands, can provide a better and hence a more accurate energy than the high rank CCSDTQ coupled cluster approach. That $E_{\RASX}$ with $\chi=10^{-4}$ is more accurate here than $E_{\rm CCSDTQ}$ follows from the fact that
$E_{\rm CCSDTQ}>E_\RASX(\chi=10^{-4})>E^0(17,2)>E_{\FCI}$.

\begin{figure}[h!]
 \includegraphics[width=0.48\textwidth]{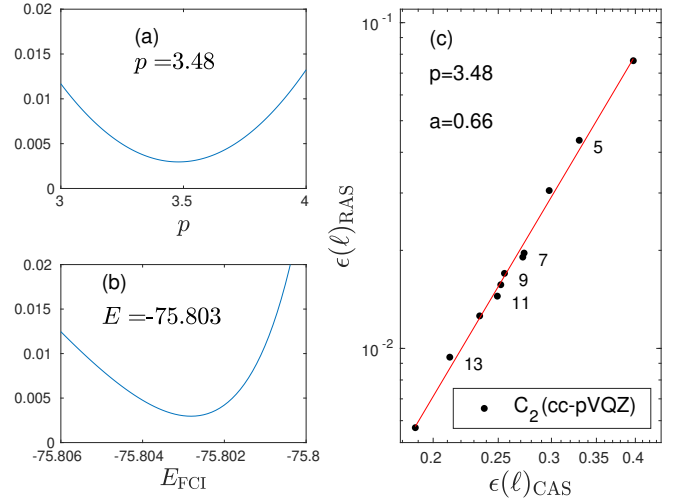}
  \caption{
Similar to Fig.~\ref{fig:cas_ras_fit_cr2}, but for the significantly larger cc-pVQZ basis set with frozen cores for $d=1.25 \AA$, corresponding to CAS(8,108). In the extrapolation procedure we have used $\ell=4\ldots 14$.
 \label{fig:cas_ras_fit_c2_108}}
\end{figure}

Finally we discuss results for the dicarbon molecule,
but for the significantly larger cc-pVQZ basis set with frozen cores for $d=1.25 \AA$ corresponding to CAS(8,108) with
$\dim {\cal H}_\FCI\approx 1\times 10^{14}$.
See Fig.~\ref{fig:cas_ras_fit_c2_108}.
Our extrapolation procedure yields $p_\RASX=3.48$ and $E_\RASX=-75.803$
using $\ell=4\ldots 14$.
Note that the error of the bare $E(14,2)=-75.7971$ is $3.6\times{10^{-3}}$ with respect to the $E_{\rm CCSD(T)}=-75.8007$ reference value.
Again the linear trend 
seen in Fig.~\ref{fig:cas_ras_fit_c2_108}(c) using $p_\RASX$ and $E_\RASX$ confirms the theory
even for very large basis sets.

\subsection{Results for a large chemical complex: FeMoco}
\label{subsec:femoco}

In this section, we apply our approach to a large chemical complex, namely the FeMoco, which is in the focus of modern quantum chemistry~\cite{Reiher-2017,Li-2019,Kai-2020,Brabec-2021}, due to its important role in nitrogen fixation (i.e., reduction of nitrogen (N$_2$) to ammonia (NH$_3$))~\cite{Hoffman-2014}, which is essential for the biosynthesis of nucleotides like DNA underlying all life forms on earth.
Computing the electronic states of the FeMoco, however, poses a great challenge and even the appropriate model space is subject to debate~\cite{Li-2019}. Here, we restrict ourselves to a model space introduced in Ref.~\cite{Reiher-2017}, since in this case there are various reference data in the literature obtained by different methods (see Ref~\cite{Kai-2020})
that we  can compare to our approach. This problem corresponds to a CAS space describing the correlation of 54 electrons on 54 orbitals, i.e. to a full Hilbert space with dimension $2.48\times 10^{31}$. 

Non-extrapolated ground state energy values obtained by DMRG and FCIQMC methods presented in  Refs.~\cite{Li-2019,Kai-2020}, as well as  our results, are summarized in Tab.~\ref{tab:femoco}.
\begin{table}[t]
  \centering
\begin{tabular}{l|r}
\hline
 \hline
 Method& Ground state energy\\
 \hline
 i-FCIQMC-RDME &-13482.17495(4)\\
 i-FCIQMC-PT2 &-13482.17845(40)\\
 sHCI-VAR & -13482.16043\\
 sHCI-PT2 & -13482.17338\\
 DMRG & -13482.17681\\
 \hline
 DMRG(D=8192)& -13482.1718\\
 DMRG(D=10240,NO)& -13482.1754\\
 RAS(23) & -13482.1421\\
 RAS(23,NO) & -13482.1544\\
 \hline
 \hline
\end{tabular}
\caption{Top: Non-extrapolated ground state energies obtained by various methods 
~\cite{Li-2019,Kai-2020}
for the FeMoco orbital space introduced in Ref.~\cite{Reiher-2017}. Bottom: our results, including data for natural orbitals as well.}
\label{tab:femoco}
\end{table}
Here, for the sake of completeness, we also performed DMRG calculations on the full orbital space for bond dimension values
up to $D=10240$. In addition, to improve the optimal basis we have performed a DMRG calculation on the full orbital space using a fixed bond dimension $D=3000$ and determined the corresponding natural orbital basis.
Since our DMRG-RAS implementation is not updated yet to handle non Abelian symmetries, we have utilized only U(1) symmetries for the full orbital space reference calculations.

\begin{figure}[h!]
 \includegraphics[width=0.48\textwidth]{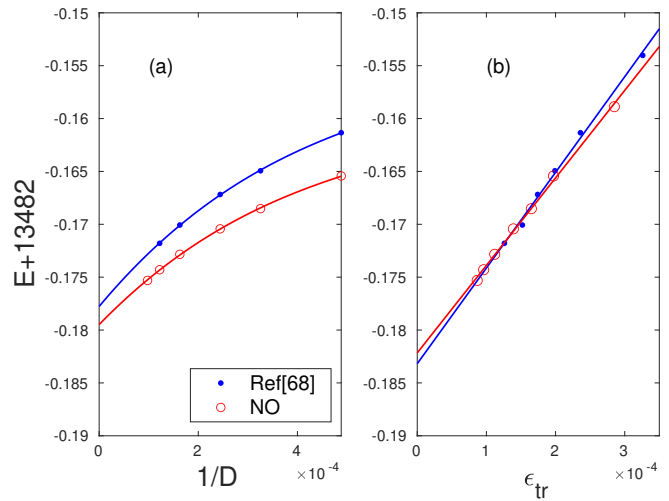}
  \caption{
(a) Ground state energy shifted by 13482 obtained by DMRG with fixed bond dimension values for the full orbital space of FeMoco, corresponding to CAS(54,54), 
for the model space taken from Ref.~\cite{Reiher-2017} and for natural orbitals. The GS energy is shown  
as a function of the inverse bond dimension. The natural orbitals were obtained by DMRG with $D=3072$. The solid line is a naive extrapolation using a 4th order polynomial.
(b) Similar, but extrapolation as a function of the truncation error, $\epsilon_{\rm tr}$, for data collected from the last DMRG sweep, 
i.e., $\min(E)$ vs $\max(\epsilon_{\rm tr})$.
}
\label{fig:femoco_54_54}
\end{figure}

Next, we discuss ground state energy values involving extrapolation. An extrapolation using large scale DMRG calculations
for the bond dimension in the range of $D=2000-4000$
provided $E=-13842.180$~\cite{Li-2019}
while extrapolation based on semi-stochastic heat bath configuration interaction (SHCI) variational and total energies gave $E\simeq-13842.1825$~\cite{Li-2019}. 
Our results for the FeMoco obtained for fixed bond dimension values on the full orbital space after ordering optimization are shown in Fig.~\ref{fig:femoco_54_54}, for two orbital sets.
In panel (a), the ground state energy shifted by 13482 
is shown for the model space taken from Ref.~\cite{Reiher-2017} 
and for natural orbitals,
as a function of the inverse bond dimension.
The solid line is a naive extrapolation using a 4th order polynomial.
It is clearly visible that for a given $D$ value lower energies are obtained for the NO-basis and 
the extrapolated energy is around $-13842.179$. 
In panel (b)   
extrapolations are presented as a function of the truncation error ($\epsilon_{\rm tr}$) for data points collected from the last DMRG sweep, i.e., $\min(E)$ vs $\max(\epsilon_{\rm tr})$.
Again, for a given $D$ value lower truncation error values are obtained for the NO basis as expected.
Nevertheless, data points fall more or less on a line and this extrapolation scheme is more reliable and more common in practice, leading to a lower ground state energy lying between $-13842.82$ and $-13842.83$. However, due to the very limited range of $\epsilon_{\rm tr}$ values a more reliable extrapolation would require even larger $D$ values. For the same reason, the more rigorous extrapolation method via the DBSS procedure by using a broader range of fixed $\chi$ \cite{Legeza-2003a} could not be used.  

\begin{figure}[h!]
 \includegraphics[width=0.48\textwidth]{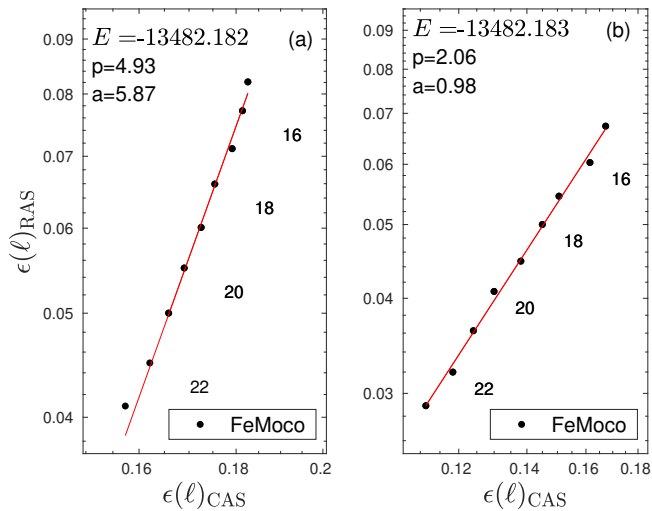}
  \caption{
(a) Result of the DMRG-RAS-X extrapolation as function of $\ell$ for the FeMoco for the model space taken from Ref.~\cite{Reiher-2017} corresponding to CAS(54,54). (b) The same but for the natural orbital basis. The predicted exponents $p_{\RASX}$, the constants $a_{\RASX}$ and the energy values $E_{\RASX}$ are also presented.
}
\label{fig:cas_ras_fit_femoco}
\end{figure}

In contrast to this, our DMRG-RAS-X method only requires significantly smaller CAS spaces.
Results using $\chi=10^{-5}$, $D_{\rm min}=2048$, $D_{\rm max}=10000$ are shown in Fig.~\ref{fig:cas_ras_fit_femoco}(a) for various $\ell$ values up to $\ell=23$. 
An almost perfect linear scaling on the log-log scale is again observed in the asymptotic regime. Here we remark that
for $\ell\ge22$, the upper bound on $D$ has been reached for a few DMRG steps; thus we omitted the last two-data points in our fit, yielding
$E=-13842.182$, $p=4.93$ and $a=5.87$.
So both our extrapolations based on $\min(E)$ vs $\max(\epsilon_{\rm tr})$ respectively DMRG-RAS-X agree to SHCI within chemical accuracy.

Repeating the same analysis for the natural orbitals,
we obtained lower $E_{\rm CAS}(\ell)$ and $E_{\rm RAS}(\ell)$ values for all $\ell$ compared to the original basis, but the enforced upper bound $M_{\rm max}=10 000$
has already been reached for $\ell\ge19$.
Therefore, in the extrapolation we have excluded the last four data points. The DMRG-RAS-X method now 
leads to a slightly lower energy for $\ell\le19$,
$E=-13842.183$, but with a much smaller exponent 
$p=2.06$. In fact, this lower exponent is 
the expected trend based on the analysis of the ladder model. Note also that both the CAS and the RAS errors
shifted to lower values due to the more optimal 
basis, as expected. 

Generating more data points within the  enforced truncation error tolerance would be desirable but requires larger bond dimensions than used here; 
nevertheless, even the excluded data points still 
lie nearly on a straight line as is shown in Fig.~\ref{fig:cas_ras_fit_femoco}(b). 

We conclude that the DMRG-RAS-X procedure has again lead to a very stable and robust extrapolation method even for a strongly correlated chemical complex like the FeMoco, requiring significantly lower computational cost than previous efforts due to smaller CAS spaces
for the model space of Ref.~\cite{Reiher-2017}.  

\section{Conclusions and outlook}
\label{sec:conclusion}

In this work, we have presented a 
promising new method for predicting the ground state energy within chemical accuracy 
for challenging strongly correlated systems and large basis sets. This was achieved by fruitfully combining three strands of research: 
1. accurate computation of core correlations using the density matrix renormalization group method and its recent advances, 
2. accounting for contributions of a large number of high energy excited configurations with small weights by the recently introduced DMRG-RAS method, with its self-consistent forward and backward coupling, 
3. a new extrapolation method based on a power law scaling of errors, which gains the final ’missing digit’ of accuracy. 
The scaling law and the extrapolation method were validated by large scale numerical simulations.

(a) We have shown that DMRG-RAS is an embedding method which accounts for forward and backward scattering between the CAS and the EXT (RAS) orbital spaces,
unlike previous post-DMRG approaches like perturbation theory or tailored coupled cluster. We also showed that the method is variational, providing an upper bound on the true energy. 

(b) We have derived a power law error scaling which relates the bare CAS energy and  the DMRG-RAS energy 
as the CAS size is varied,
by introducing and analyzing a model Hamiltonian which captures key features of the full Hamiltonian such as its two-body nature and the hierarchical structure of the CAS and RAS Hilbert spaces. Such a power law makes DMRG-RAS superior to other methods such as DMRG-TCC where errors are not even monotone.  
Based on the theory we introduced a new extrapolation method, dubbed DMRG-RAS-X, that is argued to provide the ground state energy within chemical accuracy for large systems as well. Moreover the difference between the DMRG-RAS and DMRG-RAS-X energy provides a useful intrinsic estimate of the DMRG-RAS error.

(c) We have numerically justified the validity of the power law error scaling for various small and large systems possessing single and multi reference characters, using full-CI reference energies from the literature or provided by large scale DMRG calculations on the full orbital space. We have also demonstrated that the absolute error is reduced significantly by the new extrapolation method, with better and better error reduction as the fraction of RAS orbitals increases, as inevitably happens for  large systems.

(d) We have applied DMRG-RAS-X to obtain new benchmark predictions of the full-CI ground state energy for the chromium dimer
and for the dicarbon using very large basis sets, which improve on previous high excitation rank coupled cluster reference data.

(e) We have used DMRG-RAS-X to 
obtain a carefully validated prediction of the full-CI ground state energy for the strongly correlated large chemical complex FeMoco
for the model space introduced in Ref.~\cite{Reiher-2017}.

The DMRG-RAS method is variational, free of uncontrolled errors, and -- together with the new extrapolation method, the DMRG-RAS-X -- very robust, achieving 
chemical accuracy with limited computational demands. These unique properties make
DMRG-RAS-X a vital alternative method for electronic structure calculations. 

Extension of our analysis to  ground and excited states in strongly correlated molecular clusters and to higher dimensional quantum lattice models via fermionic mode optimization~\cite{Krumnow-2016,Krumnow-2021,Mate-2022} is under progress.

\section*{Appendix: Lower bound on the FCI energy}
Here we prove the lower bound \eqref{FCIlowbd}. 
Let $H=H_0+H'$ be the partitioning of the Hamiltonian introduced in section \ref{sec:ref-ham}, with $\Delta$ as defined in \eqref{Deltadef}, and let $\Psitilde_0$ be the dressed CAS ground state (eq.~\eqref{dressed}), that is, the normalized projection of the FCI ground state onto the CAS Hilbert space $\calH_{\rm CAS}(\ell)$. The FCI ground state $ \Psi_{\rm FCI}$ can then be decomposed as
$$
  \Psi_{\rm FCI}=\sqrt{1-\delta^2}\Psitilde_0 + \Psi^\perp
$$
with inaccessible part $\Psi^{\perp}\in \calH_{\rm CAS}(\ell)^\perp$ of norm $||\Psi^\perp||=\delta\in[0,1]$. To estimate the size of $\delta$, we expand using $H=H_0+H'$ and abbreviating $b=\sqrt{1-\delta^2}$
\begin{align} \label{app1}
 & \langle\Psi_{\rm FCI}|H|\Psi_{\rm FCI}
  \rangle \\[1mm]
 &=(1-\delta^2) \underbrace{\langle \Psitilde_0|H_0|\Psitilde_0\rangle}_{=\Etilde_0} + 2\, {\rm Re} \, b\underbrace{\langle\Psitilde_0|H_0|\Psi^\perp \rangle}_{=0} 
  + \underbrace{\langle\Psi^\perp|H_0|\Psi^\perp\rangle}_{=(E_0+\Delta)\delta^2} \nonumber \\
  & + (1-\delta^2) \langle \underbrace{\Psitilde_0|H'\,|\Psitilde_0\rangle}_{=0} + 2\, {\rm Re} \, b\langle\Psitilde_0|H'\,|\Psi^\perp \rangle 
  + \underbrace{\langle\Psi^\perp|H'\,|\Psi^\perp\rangle}_{\ge 0} \nonumber  \\
  & \ge \Etilde_0 - \delta^2(\Etilde_0-E_0) + \underbrace{\Delta ||\Psi^\perp||^2 + 2 {\rm Re}\, b\langle \Psitilde_0|H'|\Psi^\perp\rangle}_{=:I(\Psi^\perp)}. \nonumber
\end{align}
Here we have used that $H'$ is $\ge 0$ on the orthogonal complement of $\calH_{\rm CAS}(\ell)$, thanks to the definition of $\Delta$ (eq.~\eqref{Deltadef}).  The auxiliary problem to minimize $I(\Phi^\perp)$ over arbitrary states $\Phi^\perp\in\calH_{\rm CAS}(\ell)^\perp$ (not required to have norm $\delta$) has the unique solution
$$
  \Phi^\perp = - \frac{1}{\Delta} \, b \, H'\Psitilde_0  
$$
(note that $H'\Psitilde_0\in\calH_{\rm CAS}(\ell)^\perp$), corresponding to the minimum value 
$$
  I(\Phi^\perp) = \underbrace{(1-2b)}_{\ge -1} \frac{1}{\Delta} ||H'\Psitilde_0||^2 \ge - \frac{1}{\Delta} ||H'\Psitilde_0||^2 = \Etilde^{(2)}.
$$
The right hand side equals $I(\Phi^\perp)$ with $b=1$, which is precisely the dressed 2nd order perturbation correction to the dressed CAS energy $\Etilde_0$. Likewise, $\Phi^\perp$ with $b=1$ is the dressed wavefunction correction $\Psitilde^{(1)}$. It follows that
\begin{equation} \label{app2}
    E_{\rm FCI} \ge \Etilde_0 + \Etilde_2 - \delta^2(\Etilde_0-E_0).
\end{equation}
We now use that the left hand side of \eqref{app1} is upper-bounded by the CAS energy $E_0$ and that the term $(1-\delta^2)\Etilde_0$ is lower-bounded by $(1-\delta^2)E_0$. It follows that
$$
 E_0\ge E_{\rm FCI} \ge E_0 + \Delta \delta^2 + 2\, {\rm Re} \, b \langle \Psitilde_0|H'|\Psi^\perp\rangle.
$$
The last term on the right hand side is, by the Cauchy-Schwarz ineqauality and the fact that $b\le 1$, lower-bounded by $-2||H'\Psitilde_0||\delta$, giving
$$
 0 \ge \Delta \delta^2 - 2 ||H'\Psitilde_0||\delta. 
$$
But the quadratic function $f(\delta)=\Delta\delta^2 - c \delta$ ($c\ge 0$) is only $\le 0$ in the interval $[0,c/\Delta]$, giving
\begin{equation} \label{app3}
   \delta \le \frac{2||H'\Psitilde_0||}{\Delta}.
\end{equation}
Substitution of \eqref{app3} into \eqref{app2} gives the final lower bound
$$
   E_{\rm FCI} \ge \Etilde_0 + \Etilde_2 - \Bigl(\frac{2||H'\Psitilde_0||}{\Delta}\Bigr)^2 (\Etilde_0-E_0). 
$$
Note that, since $H'\Psitilde_0$ belongs to $\calH_{\rm RAS}(L-\ell,2)$, this lower bound only involves Hamiltonian matrix elements and norms of states inside the accessible parts $\calH_{\rm CAS}(\ell)$ and $\calH_{\rm RAS}(L-\ell,2)$ of the full Hilbert space. 
\section*{Appendix: Stability of the scaling law for the ladder model}
In this appendix we provide a detailed derivation of the asymptotic results \eqref{app2-1}--\eqref{app2-4}. In case (a), $\Psi_0=\Psitilde_0=(1,0,0,0)^T$ and 
$$
 H' = \begin{pmatrix} 
 0 & {\rm v} & & \\
 {\rm v} & 1\! - \! \Delta & {\rm v} & \\
 & {\rm v} & 2 \!-\! \Delta & {\rm v} \\
 & & {\rm v} & 3\!-\! \Delta
 \end{pmatrix},
$$
with $\Delta$ maximal so that the bottom right $3\times 3$ block of $H'$ is $\ge$ 0. Explicitly, $\Delta=2-\sqrt{1+2{\rm v}^2} = 1 - {\rm v}^2 + O({\rm v}^4)$. Using \eqref{formulae} we find 
\begin{align}
E^{(2)} = \Etilde^{(2)} &= -{\rm v}^2 + O({\rm v}^4), \\
||\Psi^{(1)}||^2 = ||\Psitilde^{(1)}||^2 &= {\rm v}^2 + O({\rm v}^4), \\
E^{(3)} = \Etilde^{(3)} &= {\rm v}^4 + O({\rm v}^6), \\
\Etilde_0  - E_0 &= 0.
\end{align}
Consequently 
\begin{align}
  \mbox{r.h.s. of \eqref{CASlowbd} } &= \;\,{\rm v}^2 + O({\rm v}^4) \mbox{ as }{\rm v}\to 0, \\
  \mbox{r.h.s. of \eqref{RASupbd} } &= 2{\rm v}^4 + O({\rm v}^6) \mbox{ as }{\rm v}\to 0,
\end{align}
exactly matching the scaling of the actual CAS and RAS errors and hence predicting the correct exponent $p$. Note also that here the two terms $\Etilde^{(3)}$ and $||\Psitilde^{(1)}||^2\Etilde^{(2)}$ appearing in the RAS error bound are not just of  the same order as the true error, but also of same order with respect to each other, unlike what standard perturbation theory predicts!

Turning to case (c), we first determine $\Psi_0$ and $\Psitilde_0$. The first component of the CAS eigenvalue equation $H^{2\times 2}\Psi_0=E^{2\times 2}\Psi_0$ gives 
$
  E^{1\times 1}(\Psi_0)_1+{\rm v}(\Psi_0)_2 = E^{2\times 2}(\Psi_0)_1, 
$
and the first component of the full eigenvalue equation $H\Psi=E^{4\times 4}\Psi$ for the full ground state $\Psi$ gives 
$
  E^{1\times 1}(\Psitilde_0)_1+{\rm v}(\Psitilde_0)_2 = E^{4\times 4}(\Psitilde_0)_1. 
$
It follows that
$$
\Psi_0=\begin{pmatrix} 1 \\c \\0 \\ 0\end{pmatrix}\frac{1}{\sqrt{1+c^2}}, 
\;\; \Psitilde_0=\begin{pmatrix} 1 \\ \ctilde \\0 \\ 0\end{pmatrix}\frac{1}{\sqrt{1+\ctilde^2}}
$$
with $c=-(E^{1\times 1}-E^{2\times 2})/{\rm v}$,  $\ctilde=-(E^{1\times 1}-E^{4\times 4})/{\rm v}$.
Moreover, we now have
$$
 H' = \begin{pmatrix} 
 0 \; & 0 & & \\
 0 \; & 0 & {\rm v} & \\
 & {\rm v} & 2 \!-\! \Delta & {\rm v} \\
 & & {\rm v} & 3\!-\! \Delta
 \end{pmatrix},
$$
with $\Delta$ maximal so that the bottom right $2\times 2$ block of $H'$ is $\ge$ 0, i.e.,  $\Delta=\tfrac{5}{2}-\tfrac{1}{2}\sqrt{1+4{\rm v}^2} = 2 - {\rm v}^2 + O({\rm v}^4)$.
Plugging these expressions into \eqref{formulae} and using $c = -{\rm v}+{\rm v}^3+O({\rm v}^5)$, $\ctilde =-{\rm v}+ {\rm v}^3+O({\rm v}^5)$, one finds that
\begin{align}
E^{(2)} &= -\tfrac{{\rm v}^4}{2} + O({\rm v}^6), \\
\Etilde^{(2)} &= - \tfrac{{\rm v}^4}{2} + O({\rm v}^6),  \\
||\Psi^{(1)}||^2 &= \tfrac{{\rm v}^4}{4} + O({\rm v}^6), \\
||\Psitilde^{(1)}||^2 &= \tfrac{{\rm v}^4}{4} + O({\rm v}^6), \\
E^{(3)} &=\tfrac{{\rm v}^6}{4} + O({\rm v}^8), \\
\Etilde^{(3)} &= \tfrac{{\rm v}^6}{4} + O({\rm v}^8), \\
\Etilde_0  - E_0 &= O({\rm v}^6).
\end{align}
Consequently 
\begin{align}
  \mbox{r.h.s. of \eqref{CASlowbd} } &= \tfrac{{\rm v}^4}{2} + O({\rm v}^6) \mbox{ as }{\rm v}\to 0, \\
  \mbox{r.h.s. of \eqref{RASupbd} } &= \tfrac{{\rm v}^6}{4} + O({\rm v}^8) \mbox{ as }{\rm v}\to 0,
\end{align}
again matching the scaling of the actual CAS and RAS errors and predicting the correct exponent $p$.

\section*{Acknowledgments}
\"O.L. has been
supported by the Hungarian National Research, Development and Innovation Office (NKFIH) through Grant Nos.~K134983 and TKP2021-NVA-04,
by the Quantum Information National Laboratory
of Hungary, 
and by the Hans Fischer Senior Fellowship programme funded by the Technical University
of Munich – Institute for Advanced Study. 
G.B. acknowledges support from NKFIH Grant No.~FK-135496.
The development of DMRG libraries has been supported
by the Center for Scalable and Predictive methods
for Excitation and Correlated phenomena (SPEC),
funded as part of the Computational Chemical Sciences Program by the U.S. Department of Energy
(DOE), Office of Science, Office of Basic Energy Sciences, Division of Chemical Sciences, Geosciences, and Biosciences at Pacific Northwest National Laboratory.
The simulations were also performed on the national supercomputer HPE Apollo Hawk at the High Performance Computing Center Stuttgart (HLRS) under the grant number MPTNS/44246.

%\bibliographystyle{achemso}
%\bibliographystyle{abbrv}
%\bibliographystyle{apsrev4-1}
%\bibliography{references,mps_review-bibliography}{}

\end{document}